\documentclass[12pt]{article}

\usepackage{amssymb}
\usepackage[dvips]{graphicx}
\usepackage{fps}

\newcommand {\beq}{\begin{equation}}
\newcommand {\eeq}{\end{equation}}
\newcommand {\beqa}{\begin{eqnarray}}
\newcommand {\eeqa}{\end{eqnarray}}
\newcommand {\n}{\nonumber \\}
\newcommand {\Real}{\mbox{Re}}
\newcommand {\Imag}{\mbox{Im}}
\newcommand {\tr}{\mbox{tr}}
\newcommand {\trs}{\mbox{\scriptsize tr}}

\newcommand {\eff}{\mbox{\scriptsize eff}}

\newcommand {\phys}{\mbox{\scriptsize phys}}
\newcommand {\ee}{\mbox{e}}

\newcommand {\dd}{\mbox{d}}

\newcommand {\del}{\partial}

\newcommand {\new}{{\rm new}}

\font\mybb=msbm10 at 12pt
\def\bb#1{\hbox{\mybb#1}}

\def\IC{{\bb C}}

\def\IZ{{\bb Z}}

\renewcommand{\theequation}{\thesection.\arabic{equation}}

\begin{document}
\setlength{\oddsidemargin}{0cm}
\setlength{\baselineskip}{7mm}

\begin{titlepage}
 \renewcommand{\thefootnote}{\fnsymbol{footnote}}
\begin{normalsize}
\begin{flushright}
\begin{tabular}{l}
NBI-HE-00-14\\
NORDITA-2000/8-HE\\
UT-KOMABA 99-19\\
hep-th/0003208\\
March 2000
\end{tabular}
\end{flushright}
  \end{normalsize}

\vspace*{0cm}
    \begin{Large}
       \begin{center}
         {Large $N$ Dynamics of Dimensionally Reduced}\\
\vspace*{2mm}
         {4D SU($N$) Super Yang-Mills Theory} \\
       \end{center}
    \end{Large}
\vspace{2mm}

\begin{center}
J. A{\scshape mbj\o rn}$^{1)}$\footnote
            {e-mail address : ambjorn@nbi.dk},
K. N. A{\scshape nagnostopoulos}$^{2)}$\footnote
            {e-mail address : konstant@kiritsis.physics.uoc.gr},\\
W. B{\scshape ietenholz}$^{3)}$\footnote
            {e-mail address : bietenho@nordita.dk},
T. H{\scshape otta}$^{4)}$\footnote
            {e-mail address : hotta@hep1.c.u-tokyo.ac.jp}
           {\scshape and}
           J. N{\scshape ishimura}$^{1)}$\footnote{
Permanent address : Department of Physics, Nagoya University,
Nagoya 464-8602, Japan,\\
e-mail address : nisimura@nbi.dk}\\
      \vspace{5mm}
        $^{1)}$ {\itshape Niels Bohr Institute, Copenhagen University,} \\
              {\itshape Blegdamsvej 17, DK-2100 Copenhagen \O, Denmark}\\
        $^{2)}$ {\itshape Department of Physics, University of Crete,}\\
              {\itshape P.O. Box 2208, GR-71003 Heraklion, Greece}\\
        $^{3)}$ {\itshape NORDITA} \\
              {\itshape Blegdamsvej 17, DK-2100 Copenhagen \O, Denmark} \\
        $^{4)}$ {\it Institute of Physics, University of Tokyo,}\\
                 {\it Komaba, Meguro-ku, Tokyo 153-8902, Japan}
\end{center}


\vspace{5mm}


\begin{abstract}
\noindent
We perform Monte Carlo simulations of a supersymmetric matrix model, 
which is obtained by dimensional reduction
of 4D SU($N$) super Yang-Mills theory.
The model can be considered as a four-dimensional counterpart of
the IIB matrix model.
We extract the space-time structure represented by the 
eigenvalues of bosonic matrices. In particular we
compare the large $N$ behavior of the space-time extent
with the result obtained from a low energy effective theory.
We measure various Wilson loop correlators which represent
string amplitudes and we observe a nontrivial universal scaling in $N$.
We also observe that the Eguchi-Kawai equivalence to
ordinary gauge theory does hold at least within a finite range of scale. 
Comparison with the results for the bosonic case clarifies
the r\^{o}le of supersymmetry in the large $N$ dynamics.
It does affect the multi-point correlators qualitatively,
but the Eguchi-Kawai equivalence is observed even in the bosonic case.
\end{abstract}
\vfill
\end{titlepage}
\vfil\eject

\setcounter{footnote}{0}
\section{Introduction}
\setcounter{equation}{0}

\renewcommand{\thefootnote}{\arabic{footnote}} 
A recent excitement in string theory is
that we finally arrive at concrete proposals 
for nonperturbative definitions of superstring theory 
\cite{BFSS,IKKT,nonperturbative,nonperturbative2}.
In particular, Matrix Theory \cite{BFSS} 
and the IIB matrix model \cite{IKKT}, 
which are candidates for a nonperturbative definition of
M-theory and type IIB superstring theory, respectively, 
have attracted considerable interest (for reviews, 
see Refs.\ \cite{Mrev,IIBrev}).
The proposed formulations
take the form of large $N$ reduced models \cite{EK},
which can be obtained by dimensional reduction 
of large $N$ gauge theories;  a reduction to one dimension for M-theory,
and to zero dimension (one point) for type IIB superstring theory.
These proposals are supported by some evidences
such as the similarity of the Hamiltonian (or the action)
to that of membranes or strings, the appearance of 
soliton-type objects known as D-branes with consistent interactions,
and the consistency with string dualities upon compactification.
For the IIB matrix model,
even an attempt to establish a direct connection 
to perturbative string theory has been made
by deriving the light-cone string field Hamiltonian
from loop equations of the model \cite{FKKT}. This attempt was indeed
successful, albeit with the aid of symmetry and power-counting arguments.
Further variants of that model have been proposed in Refs.\ \cite{variants}.

As another approach to analyze these proposals we can
investigate the dynamical properties of 
large $N$ reduced models of this kind,
and verify if they really have the potential to describe
nonperturbative string theory.
In Ref.\ \cite{2DEK} a two-dimensional reduced model with unitary matrices
has been studied in this context. There, a large $N$ limit, which
differs from the planar limit (or 't Hooft limit), has been
discovered numerically
\footnote{This non-planar large $N$ limit has recently been re-interpreted 
as a continuum limit of non-commutative gauge theory \cite{AMNS}.}.
A Hermitian matrix model obtained by simply omitting the fermions
in the IIB matrix model, and its generalizations to arbitrary dimensions
larger than two, have been studied in Refs.\ \cite{KS,HNT}.
In Ref.\ \cite{HNT}, Monte Carlo simulations up to $N=256$ have been reported
and analytical methods such as perturbation theory, 
Schwinger-Dyson equations and $1/D$ expansions have been applied,
providing a comprehensive understanding of the 
large $N$ dynamics of that model.
A new type of Monte Carlo technique
was used to extract the value of the partition function \cite{KNS,KS}.
This technique has been further applied to extract 
the asymptotic behavior of the eigenvalue distribution 
for large eigenvalues \cite{Eigen,Potsdam}.

In the present paper, we make a first attempt to extract
the large $N$ dynamics of a {\em supersymmetric large $N$ reduced model}
obtained by dimensional reduction of 4D
supersymmetric Yang-Mills theory.
The model can be regarded as a 4D counterpart of
the IIB matrix model. The bosonic model
is well understood \cite{KS,HNT}, but the
inclusion of fermions makes the system far more complicated.
An attempt to study the model analytically
maps it onto a soluble system \cite{analytic},
but the relevance to the original model is unclear
due to a nontrivial change of variables in an analytic continuation.
Here we take a direct approach, based on Monte Carlo simulations.
Fermions are completely included by the use of the
so-called Hybrid-R algorithm \cite{hybridR}, 
which is one of the standard methods in QCD simulations 
with dynamical quarks.

One of the features that makes the IIB matrix model 
most attractive as a nonperturbative definition of string theory
is that space-time is dynamically generated
as the eigenvalue distribution 
of the bosonic matrices \cite{IKKT,AIKKT,IK}.
In Ref.\ \cite{AIKKT} a low energy effective theory of the model
is constructed, 
where the authors discuss some possible mechanisms that
may induce a collapse of the eigenvalue distribution
to a four-dimensional manifold.
We extract the large $N$ behavior of 
the space-time extent in our model and
compare the result with the prediction obtained by the low
energy effective theory.
Another dynamical issue to be addressed 
in this context
is the space-time uncertainty relation,
which was proposed as a principle for constructing 
nonperturbative string theory \cite{Yoneya}.
We extract the large $N$ behavior of 
the space-time uncertainty of our model
and confirm that the model indeed satisfies the proposed principle.

Another attractive feature of the IIB matrix model
as a nonperturbative definition of string theory
is that its only parameter $g$ is a simple scale 
parameter \footnote{This means in particular that the string coupling
constant, which is related to the vacuum expectation value of 
the dilaton field, 
is {\em not} a tunable parameter.
We come back to this point in Section \ref{interpretation}.}.
One has to tune $g$ suitably as one sends $N$ to infinity,
so that the correlation functions have finite large $N$ limits.
According to Ref.\ \cite{FKKT}, Wilson loop operators can be interpreted as
the string creation and annihilation operators, and
it was found that $g^2 N$ should be fixed in order to obtain
the light-cone string field Hamiltonian in the large $N$ limit.
It is a non-trivial test of the model to verify
if the correlation functions of Wilson loops 
really have a universal large $N$ scaling.
We address this issue in the present model and show that
there is indeed a universal large $N$ scaling at fixed $g^2 N$.

We also address yet another important dynamical issue 
in this model, namely the question of equivalence 
to ordinary super Yang-Mills theory
in the sense of Eguchi and Kawai \cite{EK},
which is exactly the way large $N$ reduced models first appeared
in history. The crucial observation is
that large $N$ gauge theory does not depend on the volume
(under some assumptions), which inspired Eguchi and Kawai to propose
the zero-volume limit of large $N$ gauge theory as
a model equivalent to the gauge theories in an infinite volume \cite{EK}.
One of the assumptions is that the $(\IZ_N )^D$ symmetry
of the model
is {\em not} spontaneously broken, where $D$ is the space-time dimension.
However, in the purely bosonic case 
in $D>2$, the symmetry {\em is} spontaneously broken at weak coupling \cite{BHN},
thus preventing one from taking a continuum limit.
This led to modifications of the model \cite{BHN,Parisi,GK,DW,AS}
so that the $(\IZ_N )^D$ symmetry
is not spontaneously broken while keeping the equivalence valid.
In the supersymmetric case, 
the effective action which induces the spontaneous symmetry breaking of the
(\IZ$_N$)$^D$ symmetry
is naively cancelled by the contributions of fermions.
Indeed, in the scalar field case, it has been shown that the reduced model
is equivalent to the field theory without such modifications \cite{MK}.
We observe in the present supersymmetric model that
the Eguchi-Kawai equivalence indeed holds at least
in a finite range of scale.
What is rather remarkable is that actually this is true also for
the bosonic case, which is contrary to what has been expected.

In Section \ref{model} we describe the model we are going to investigate.
In Section \ref{spacetime} we study the space-time structure
of the model. 
In Section \ref{wilsonloops} we present our results for
 correlation functions of Wilson loop and Polyakov line operators,
and we discuss 
the Eguchi-Kawai equivalence as well as the universal scaling behavior.
Section \ref{summary} is devoted to a summary and discussion.
In Appendix \ref{algorithm} we comment on the algorithm 
we used for the simulation.
In Appendix \ref{bosonic_results} we present the corresponding
results for the bosonic case for comparison.

\section{The model}
\setcounter{equation}{0}
\label{model}

The model we investigate is a supersymmetric matrix model
obtained by dimensional reduction of 4D
SU($N$) super Yang-Mills theory.
The partition function is given by
\begin{eqnarray}
Z &=& \int \dd A ~  \ee ^{-S_b} \int \dd \psi \dd \bar{\psi}
~ \ee ^{- S_f } , \nonumber \\
S_b &=& -\frac{1}{4 g^2}  \tr [A_{\mu},A_{\nu}]^{2} , \nonumber \\
S_f  &=& - \frac{1}{g^2}
 \tr \Big( \bar{\psi}_\alpha  (\Gamma^{\mu})_{\alpha\beta} 
[A_{\mu},\psi _\beta] \Big) ,
\label{action}
\end{eqnarray}
where $A_\mu$ ($\mu=1 ,  \dots , 4$) are 
traceless $N \times N$ Hermitian matrices, and 
$\psi_\alpha$, $\bar{\psi}_\alpha$ ($\alpha = 1,2$) 
are traceless $N \times N$ complex matrices.
The measure is defined as
\beqa
\dd \psi \dd \bar{\psi}  &=&  
\prod_{\alpha = 1 }^2 
\left[ \prod_{i,j=1}^N 
\Big[ \dd (\psi _\alpha)_{ij} 
\dd (\bar{\psi}  _\alpha )_{ij} \Big] \
 \delta \Big( \sum _{i=1} ^N (\psi _\alpha)_{ii} \Big) 
\delta \Big( \sum _{i=1} ^N ( \bar{\psi} _\alpha )_{ii}  \Big)
\right] ,  
\label{measure_fermion}
 \\
\dd A  &=&  \prod_{\mu = 1 }^4
\left[ \prod_{i<j} \{\dd \Real (A_\mu)_{ij} \dd 
\Imag (A_\mu)_{ij} \}  \prod_{i=1} ^N \{ \dd (A_\mu)_{ii} \}
\delta \Big( \sum _{i=1} ^N (A_\mu)_{ii} \Big) \right] .
\eeqa
This model is invariant under 
4D Lorentz transformations 
\footnote{When one defines the IIB matrix model
  nonperturbatively, a Wick rotation to
  Euclidean signature in needed. This is also the case for the present
  model. Hence by Lorentz invariance we actually mean
  rotational invariance.},
where $A_{\mu}$ transforms as a vector and
$\psi _ \alpha$ as a Weyl spinor.
$\Gamma_\mu$ are 2 $\times$ 2 matrices acting on the spinor
indices, and they can be given explicitly as
\begin{equation}
\Gamma_1 = i  \sigma_1 , \
\Gamma_2 = i \sigma_2 , \
\Gamma_3 = i  \sigma_3 , \
\Gamma_4 =   {\bf 1} .
\label{Gamma}
\end{equation}
The model is manifestly supersymmetric,
and it also has a SU($N$) symmetry
\beq
\label{SU_N}
A_\mu  \rightarrow  V A_\mu V^\dag  ~~~;~~~
\psi _\alpha \rightarrow V \psi _\alpha V^\dag   ~~~;~~~
\bar{\psi} _\alpha \rightarrow V \bar{\psi} _\alpha V^\dag   ,
\eeq
where $V\in \mbox{SU}(N)$.
All these symmetries are inherited from the super Yang-Mills theory
before dimensional reduction.
The model can be regarded as the four-dimensional counterpart of
the IIB matrix model \cite{IKKT}.

In contrast to unitary matrix models,
where the integration domain for the partition function is compact,
the first nontrivial question to be addressed
in Hermitian matrix models in general,
is whether the model is well-defined as it stands.
The problem can be most clearly understood 
by decomposing the Hermitian matrices
into eigenvalues and angular variables,
where a potential danger of divergence exists 
in the integration over the eigenvalues, even at finite $N$.
This issue has been addressed numerically
for the supersymmetric case \cite{KNS} at $N=3$
as well as the bosonic case \cite{KS} up to $N=6$.
Exact results are available for $N=2$ \cite{SuyamaTsuchiya}.
There is also a perturbative argument which is valid when
all the eigenvalues are well separated from each other
\cite{HNT,AIKKT}. This reasoning agrees with the conclusions obtained
for small $N$ \cite{KS,KNS,SuyamaTsuchiya}.
In particular, supersymmetric models in $D=4,6,10$
are expected to be well-defined for arbitrary $N$.
Our simulations confirm that this is indeed the case for $D=4$.

Since the model is well-defined without any cutoff, 
the parameter $g$, which is the only parameter of the 
model, can be absorbed by rescaling the variables,
\beqa
\label{rescaleA}
A_\mu &=& 
g^{1/2} X _\mu  \ ,\\
\psi_\alpha &
=& g^{3/4} \Psi _\alpha  \ .
\eeqa
Therefore, $g$ is a scale parameter rather than
a coupling constant, i.e.\
the $g$ dependence of physical quantities is completely 
determined on dimensional grounds. The parameter
$g$ should be tuned appropriately as one sends $N$ to infinity,
so that each correlation function of Wilson loops
has a finite large $N$ limit.
Whether such a limit really exists or not is 
one of the dynamical issues we address in this work.

We now discuss the Eguchi-Kawai equivalence \cite{EK},
which is the equivalence between reduced models and 
the corresponding gauge theories
in the large $N$ limit.
In its proof based on the Schwinger-Dyson equation,
one has to assume quantities of the type
\beq
\langle \tr (\ee ^{i k_\mu A_\mu}) \tr (\ee ^{-i k_\mu A_\mu})
\rangle ~~~~~~~(k_\mu \neq 0)
\label{twopoint}
\eeq
to vanish.
Assuming in addition large $N$ factorization, 
the vanishing of (\ref{twopoint}) is equivalent to
$ \langle \tr (\ee ^{i k_\mu A_\mu}) \rangle$$= 0$,
which is guaranteed if the eigenvalues
of $A_\mu$ are uniformly distributed on the whole real axis in the
large $N$ limit.
The fact that the present model is well-defined 
without any cutoff implies that the eigenvalue distribution
of $A_\mu$ is {\em not} uniform, but it has a finite extent for finite $N$.
Hence, the Eguchi-Kawai equivalence
is quite nontrivial
even in the supersymmetric case.
Here the situation is more subtle than in the case of
the unitary matrix model version of a large $N$ reduced model \cite{EK}.
There, the model has the $(\IZ_N )^D$ symmetry
$U_{\mu} \rightarrow 
\ee ^{2 \pi i m_{\mu}/N} U_{\mu}$ \ ($m_\mu=0,1,\cdots,N-1$),
hence quantities like $\langle \tr (U_\mu)^n \rangle$
vanish, unless the symmetry is spontaneously broken.

One might be tempted to consider a model 
defined by the partition function (\ref{action})
but without imposing the traceless condition on $A_\mu$.
We denote such a model as the U($N$) model, to be distinguished from
the original model, which we call the SU($N$) model.
The U($N$) model has the U(1)$^4$ symmetry
\beq
A_\mu \rightarrow  A_\mu + \alpha _\mu  {\bf 1}_N ,
\label{U1_4sym}
\eeq
where $\alpha _\mu$ is a real vector.
Note, however, that the trace part of $A_\mu$
in the U($N$) model simply decouples because
$A_\mu$ appears in the action only through commutators.
The transformation (\ref{U1_4sym}) acts on the decoupled trace
part
and hence it cannot play any physical r\^{o}le.
Indeed the quantity (\ref{twopoint}) calculated with the U($N$) model 
or with the SU($N$) model is exactly the same.
Thus, considering the U($N$) model does not help.

Next we comment on the method we use to study the model.
Details can be found in Appendix A.
The integration over fermionic variables can be done explicitly
and the result is given by \ $\det {\cal M} $,
${\cal M}$ being a
$2(N^2-1)$ $\times$ $2(N^2-1)$ complex matrix which depends
on $A_\mu$.
Hence the system we want to simulate can be written
in terms of bosonic variables as
\beq
Z =  
\int \dd A  ~ \ee ^{-S_b} \det {\cal M} \ .
\eeq
A crucial point for the present work is that
the determinant \ $\det {\cal M}$ \ is actually
real positive, as we prove in Appendix A.
Due to this property,
we can introduce a $2(N^2-1)$ $\times$ $2(N^2-1)$
Hermitian positive matrix ${\cal D}= {\cal M}^\dag {\cal M}$,
so that \ $\det {\cal M} = \sqrt{\det {\cal D}}$, and
the effective action of the system takes the form
\beq  \label{eff_act}
S_{\eff} = S_b -  \frac{1}{2} \ln \det {\cal D} \ .
\eeq
We apply the Hybrid R algorithm \cite{hybridR} to simulate
this system.
In the framework of this algorithm, 
each update of a configuration is made by
solving a Hamiltonian equation for a fixed ``time'' $\tau$.
The algorithm is plagued by a systematic error due to the 
discretization of $\tau$ that we used to solve the equation numerically.
We performed simulations at three different values of 
the time step $\Delta \tau$.
Except in Fig.\ \ref{fig:Rnew}, 
we find that the results do not depend much
on $\Delta \tau$ (below a certain threshold),
so we just present the results for the value 
$\Delta \tau = 0.002$, which appears to be sufficiently small.

We also note that there is an exact result
\beq \label{exact}
\langle \tr F^{2} \rangle =
- \langle \tr (\sum_{\mu \neq \nu} [A_\mu , A_\nu] ^2) \rangle
= 6 g^2 (N^2 - 1 ) ,
\label{trF2}
\eeq
which can be obtained by a scaling argument, similar to the
one used for the bosonic case \cite{HNT}.
We used this exact result
to check the code and the numerical accuracy.

\section{The space-time structure}
\setcounter{equation}{0}
\label{spacetime}

We first study the space-time structure of the reduced model.
In the IIB matrix model, the eigenvalues of the bosonic matrices
$A_\mu$ are interpreted 
as the space-time coordinates \cite{IKKT,IIBrev,AIKKT,IK}.
However, since the matrices
$A_\mu$ are not simultaneously diagonizable in general,
the space-time is not classical.
In order to extract the space-time structure, 
we 
first define the space-time uncertainty $\Delta$ by
\beq
\Delta ^2  = \frac{1}{N} \tr (A_\mu^{~2})
- \max_{U \in \mbox{\scriptsize SU}(N)} \frac{1}{N}
\sum_i \{ (U A_\mu U^\dagger)_{ii}  \} ^2 ,
\label{maxU}
\eeq
which is invariant under Lorentz transformation
and SU($N$) transformation (\ref{SU_N}) \cite{HNT}.
This formula has been derived in Ref.\ \cite{HNT}
based on analogy to quantum mechanics,
regarding $A_\mu$ as an operator acting on a space of states.
It has the natural property that
$\Delta ^2$ vanishes if and only if the matrices $A_\mu$ are 
diagonalizable simultaneously.
For each configuration $A_\mu$ generated by a Monte Carlo simulation, 
we maximize 
$\sum_i \{ (U A_\mu U^\dagger)_{ii}  \} ^2 $ with respect to 
the SU($N$) matrix $U$.
We denote the matrix which yields the maximum as $U_{\max}$,
and we define $x_{\mu i} = (U_{\max} A_\mu U_{\max}^\dagger)_{ii}$ as the
space-time coordinates of $N$ points ($i=1,\cdots ,  N$)
in four-dimensional space-time.

Note that
$x_{\mu i}$ should be identified with the dynamical variables
denoted by the same $x_{\mu i}$ in Ref. \cite{AIKKT}.
There, the bosonic matrices 
$A_\mu$ and the fermionic matrices $\psi _\alpha$
are decomposed into diagonal and off-diagonal elements as
\beqa
(A_\mu)_{ij} &=& x_{\mu i} \delta_{ij} + a_{\mu ij} ~~~~~(a_{\mu ii}=0),\n
(\psi_\alpha)_{ij} &=& \xi_{\alpha i} \delta_{ij} 
+ \varphi_{\alpha ij}  ~~~~~(\varphi_{\alpha ii}=0) .
\label{decompose}
\eeqa
The off-diagonal parts $a_{\mu ij}$ and $\varphi_{\alpha ij}$
are integrated out using the ``Lorentz gauge'' in the one-loop 
approximation, which is valid when the points $x_{\mu i}$ $(i=1,\cdots,N)$ 
are well separated from each other.
Thus one obtains the effective action for $x_{\mu i}$ and $\xi_{\alpha i}$,
which can be considered as a low-energy effective action of the 
supersymmetric large $N$ reduced model.
In order to get the effective action only for $x_{\mu i}$,
one still has to integrate over $\xi_{\alpha i}$,
which cannot be done exactly for $D=6$ and $D=10$ (IIB matrix model).
In $D=4$, however, the integration over $\xi_{\alpha i}$ can be
carried out exactly and the system of $x_{\mu i}$ 
is described by a simple branched polymer
with an attractive potential between the points connected by a bond.
In $D=6$ and $D=10$, the system of $x_{\mu i}$ is expected to be
described by some complicated branched-polymer like structure.
Thus, although the one-loop approximation might seem quite
drastic, the low energy effective theory of $x_{\mu i}$
still has a nontrivial dynamics.
In Ref.\ \cite{AIKKT}, some plausible mechanisms
for the collapse of the $x_{\mu i}$ distribution
in IIB matrix model have been discussed.
What we have described in the previous paragraph provides
a way to extract the low-energy effective theory
of $x_{\mu i}$ from the full model without perturbative expansions.
In particular, we can check explicitly whether the one-loop approximation
adopted in Ref. \cite{AIKKT} really
captures the low energy dynamics of the 
supersymmetric large $N$ reduced model.

\begin{figure}[hbt]
\begin{center}
\hspace{1cm}
    \includegraphics[height=9cm]{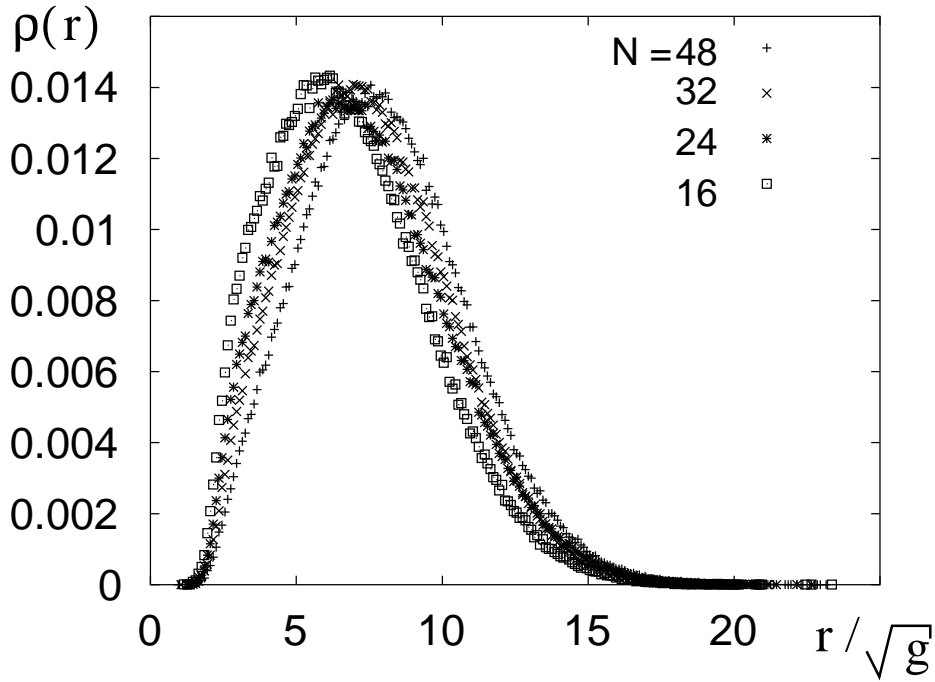}
\end{center}
    \caption{The distribution of distances $\rho(r)$,
     plotted against $r/\protect\sqrt{g}$ for $N=16$, 24, 32 and 48.}
\label{fig:EV04}
\end{figure}

We first look at the distribution $\rho (r) $ of the distances $r$,
where the distance between two arbitrary points $x_i \neq x_j$
is measured by $\sqrt{(x_i - x_j)^2}$.
In Fig.\ \ref{fig:EV04} we plot the results 
for $N=16,\, 24,\, 32$ and 48.
We first note that the distribution at small $r$ falls off
rapidly below $r/\sqrt{g}\sim 1.5$, independently of $N$.
(This behavior is also seen in the bosonic case shown in 
Fig.\ \ref{fig:BEV1}.)
This observation is in agreement with the argument in Ref.\ \cite{AIKKT} 
that the ultraviolet behavior of the space-time structure of the model
is controlled by the SU(2) matrix model.
There, this argument has been used to justify the introduction of 
a $N$-independent ultraviolet cutoff in the low energy effective theory,
which otherwise suffers from ultraviolet divergence due to coinciding
$x_{\mu i}$'s. Our observation confirms that 
the ultraviolet cutoff is indeed generated dynamically 
if one treats the full model nonperturbatively 
instead of making perturbative expansions around diagonal matrices.

\begin{figure}[htbp]
  \begin{center}
    \includegraphics[height=12cm,angle=270]{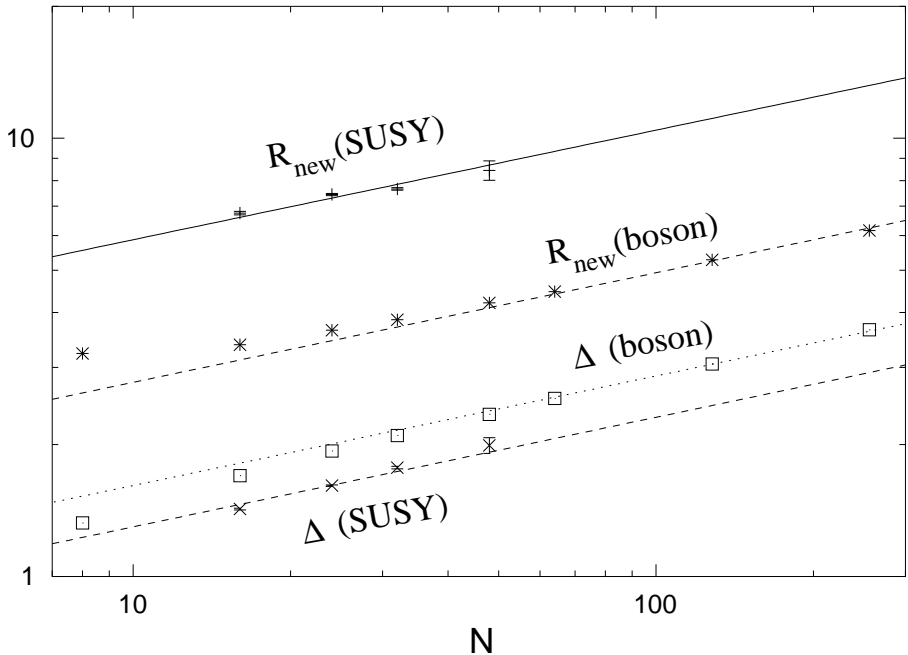}
\hspace{1cm}
    \caption{ $R_{\new}/\protect\sqrt{g}$ and 
$\protect\sqrt{\langle \Delta ^2 \rangle / g }$, 
plotted against $N$.
The results for the bosonic model are also included.
The lines are fits to the power behavior $\propto N^{1/4}$,
which is predicted theoretically. 
(In the labels we use a short-hand notation.)}
\label{fig:Rnew}
  \end{center}
\end{figure}

In both, the supersymmetric as well as the bosonic case,
we observe that the distribution shifts towards larger $r$
as one increases $N$.
In order to quantify this behavior,
we define the extent of space-time by
\beq
R _{\new} = \frac{2}{N(N-1)} 
\left\langle \sum_{i < j}  \sqrt{(x_i - x_j)^2}   \right\rangle 
= \int_0 ^\infty  \dd r ~ r \rho (r) .
\label{Rnewdef}
\eeq
We denote this quantity by $R _{\new}$ in order to distinguish
it from the definition of the extent of the space-time
$R= \sqrt{\langle \frac{1}{N}
\tr (A_\mu^{~2}) \rangle}$ 
used in Ref.\ \cite{HNT}.
$R$, which roughly corresponds to 
$\sqrt{\langle \int_{0}^{\infty} \dd r \, r ^2 \rho (r)\rangle}$,
is logarithmically divergent in the 4D supersymmetric case
due to the asymptotic behavior $\rho (r) \sim r^{-3}$ at large $r$ 
\cite{Eigen}.
On the other hand,
$R_{\new}$ does not suffer from this divergence
as eq.\ (\ref{Rnewdef}) shows. In Fig.\ \ref{fig:Rnew} 
we plot the results for the space-time extent $R_{\new}$ 
as well as those for the space-time uncertainty 
$\sqrt{\langle \Delta ^2 \rangle}$ 
for $N=16,24,32$ and 48.
We repeat the same measurements for the bosonic model 
with $N$ up to 256 and include
the results in Fig.\ \ref{fig:Rnew} for comparison.
We see that the effect of fermions enhances $R_{\new}$ and 
suppresses $\sqrt{\langle \Delta ^2 \rangle}$ considerably.
However, the power of the large $N$ behavior does not seem
to be affected.

Let us discuss the results for $R_{\new}$.
In the bosonic case, the data can be nicely fitted to a power behavior 
with $R_{\new}/\sqrt{g} = 1.56(1) \cdot   N^{1/4}$.
%
As expected, the observed large $N$ behavior of $R_{\new}$ is
the same as the one obtained for $R$ in Ref. \cite{HNT}.
In the supersymmetric case,
the large $N$ behavior of $R_{\new}$ can be predicted
by the branched polymer picture based on the
one-loop approximation \cite{AIKKT}.
Since the Hausdorff dimension of branched polymers
is four, $d_{\mbox{\scriptsize H}}=4$, the number of points $N$ grows as 
the extent $R_{\new}$ of the branched polymer,
$N \sim (R_{\new} /\ell) ^{d_{\mbox{\tiny H}}}$. Here
$\ell$ is the minimum length of the bond, which is of $O(\sqrt{g})$
as we have already discussed.
Thus one obtains $R_{\new}\sim \sqrt{g} \, N^{1/4}$.
The data in Fig.\ \ref{fig:Rnew} seem to be consistent with this prediction.
Fitting the data to this power behavior, we obtain
$R_{\new}/\sqrt{g} = 3.30(1) \cdot N^{1/4}$.

One might be surprised that supersymmetry does not affect
the power of the large $N$ behavior of the space-time extent $R_{\new}$.
We recall, however, that in the bosonic case 
the explanation is completely different 
--- although the power is the same \cite{HNT}.
There the one-loop perturbative expansion 
around diagonal matrices yields
a logarithmic attractive potential between all the pairs of
eigenvalues.
The one-loop effective potential is dominant as far as the extent 
of the eigenvalue distribution is larger than $\sqrt{g} \, N^{1/4}$.
One can therefore put an upper bound on the space-time extent 
$R \lesssim \sqrt{g} \, N ^{1/4}$.
What happens actually is that this upper bound is saturated.
The behavior $R \sim \sqrt{g} \, N ^{1/4}$ can also be shown to all orders
in the $1/D$ expansion \cite{HNT}.

Let us turn to the results for the space-time uncertainty.
Using the one-loop perturbative expansion,
$\langle \Delta ^2 \rangle$ can be roughly estimated as
\beq
\langle \Delta ^2 \rangle = \frac{1}{N} \sum _{ij} 
\langle a_{\mu ij} a_{\mu ji} \rangle 
= \frac{1}{N} \sum_{ij} \left\langle  \frac{g^2}{(x_i - x_j)^2} 
\right\rangle
\sim \frac{g^2 N}{R_{\new}^2} \ .
\eeq
The powers of $R_{\new}$ and $\sqrt{\langle \Delta ^2 \rangle}$, 
as well as the coefficients
we observe, are in qualitative agreement with this estimation.
The bosonic case has been studied before in Ref.\ \cite{HNT}. The data 
in Fig.\ \ref{fig:Rnew} can be nicely fitted to a power behavior with
$\sqrt{\langle \Delta ^2 \rangle / g } = 0.907(3) \cdot  N^{1/4}$.
Thus, in the bosonic case we obtain $\sqrt{\langle \Delta ^2 \rangle}
\sim 0.58 \, R_{\new}$ \cite{HNT}, 
which indicates a signficant deviation from the classical 
space-time picture. 
On the other hand, in the supersymmetric case we obtain
$\sqrt{\langle \Delta ^2 \rangle / g} = 0.730(2) \cdot N ^{1/4}$,
hence our result amounts to 
$\sqrt{\langle \Delta ^2 \rangle}\sim 0.22 \, R_{\new}$, coming
closer to the classical space-time picture.

We will see in the next section that 
the scale parameter $g$ should be taken to be
$O(1/\sqrt{N})$ in order to obtain
a universal scaling behavior for the Wilson loop correlators.
This means that the space-time uncertainty
in the physical scale remains finite, rather than vanishing, 
in the large $N$ limit.
Therefore the present model satisfies the space-time
uncertainty principle proposed for nonperturbative definitions
of string theories \cite{Yoneya}.

\section{Wilson loop correlation functions}
\setcounter{equation}{0}
\label{wilsonloops}

In the interpretation of a large $N$ reduced model as a string theory,
Wilson loop operators correspond to string creation operators
\cite{FKKT}.
Therefore, the existence of a non-trivial large $N$ limit of the
Wilson loop correlators is an absolutely crucial issue.
It has been addressed before in the 2D Eguchi-Kawai model,
where non-trivial large $N$ scaling has indeed been observed \cite{2DEK}.

We define the ``Wilson loop'' and the
``Polyakov line'' operators as
\beqa
W(k) &=& \frac{1}{N} \tr ( \ee ^{i k X_1}
\ee ^{i k X_2} \ee ^{-i k X_1} \ee ^{-i k X_2}) \ , \n
P (k) &=& \frac{1}{N} \tr ( \ee ^{i k X_1}) \ ,
\label{wilsondef}
\eeqa
where $X_\mu$ are dimensionless matrices
defined in eq.\ (\ref{rescaleA}).
For convenience we have chosen particular components of $X_\mu$
in the above definitions, but the choice of the directions becomes 
irrelevant when taking the vacuum expectation value,
due to Lorentz symmetry and parity invariance.
In the actual calculations
we take an average over all possible choices
of the components in order to enhance the statistics.

The real parameter $k$ represents the dimensionless ``momentum'' 
that characterizes the momentum density distributed along the string.
The physical (dimensionful) momentum variable is given by
$k_{\phys} = k/\sqrt{g}$.
We have to tune $g$ depending on $N$, so that
the correlation functions of the above operators
have definite large $N$ limits as functions of $k_{\phys}$.
In the following, we always set $g=1$ for $N=48$ 
without loss of generality.

In all plots except for Fig.\ \ref{fig:Wilson1-log},
we further assume $g$ to be proportional to $1/\sqrt{N}$. This turns out
to be consistent with large $N$ scaling, hence $g \propto 1/\sqrt{N}$
can be regarded as one of our observations.

\subsection{One-point function and Eguchi-Kawai equivalence}
\label{onepoint}

In this subsection we discuss the one-point functions, and
we start with the Wilson loop $\langle W(k) \rangle$.
Also Ref.\ \cite{Potsdam} presents some recent results about
this quantity.

\begin{figure}[htbp]
  \begin{center}
    \includegraphics[height=12cm,angle=270]{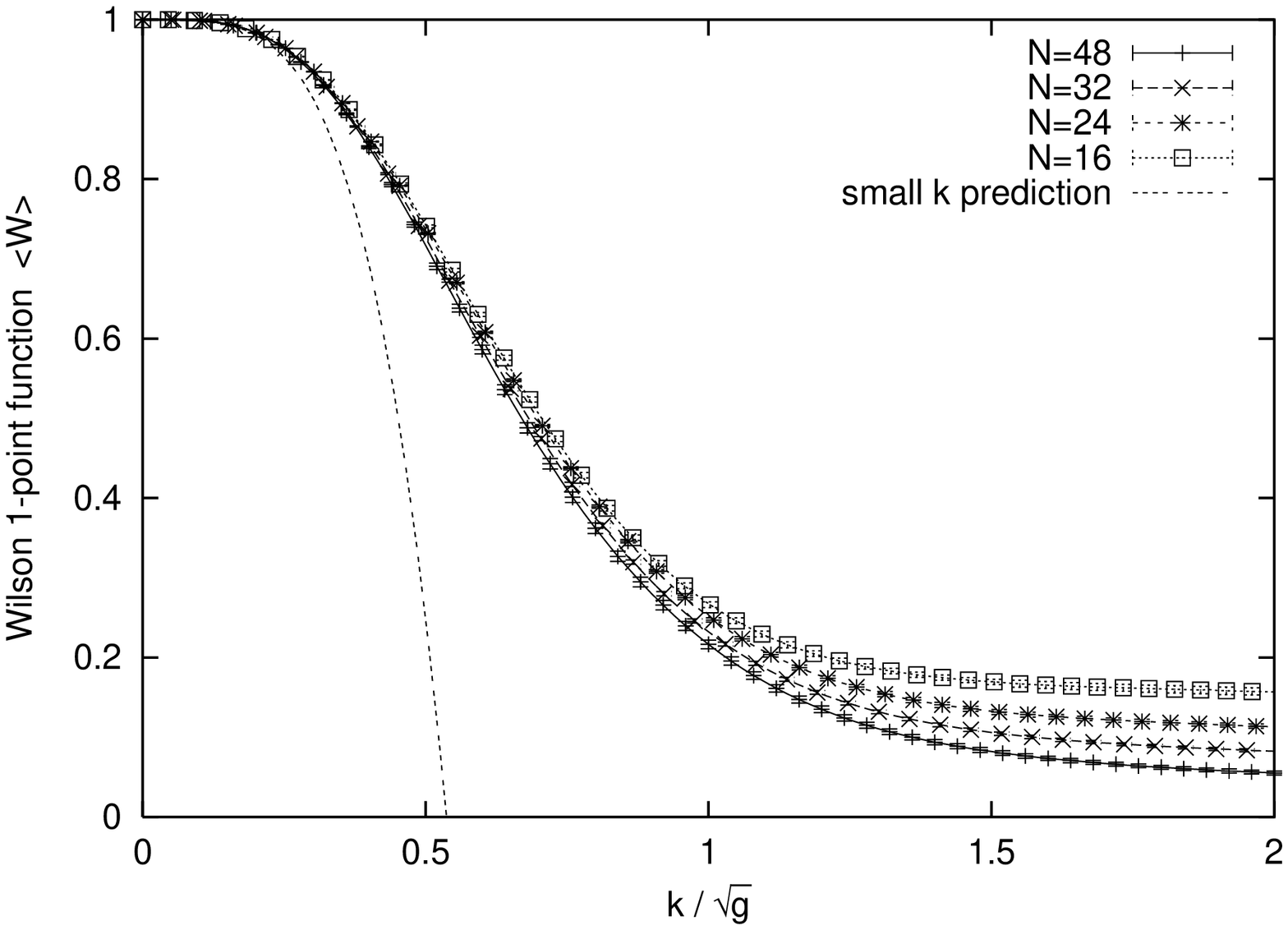}
    \caption{The Wilson 1-point function $\langle W \rangle$,
plotted against $k_{\phys} = k/\protect\sqrt{g}$.}
\label{fig:Wilson1}
    \includegraphics[height=12cm,angle=270]{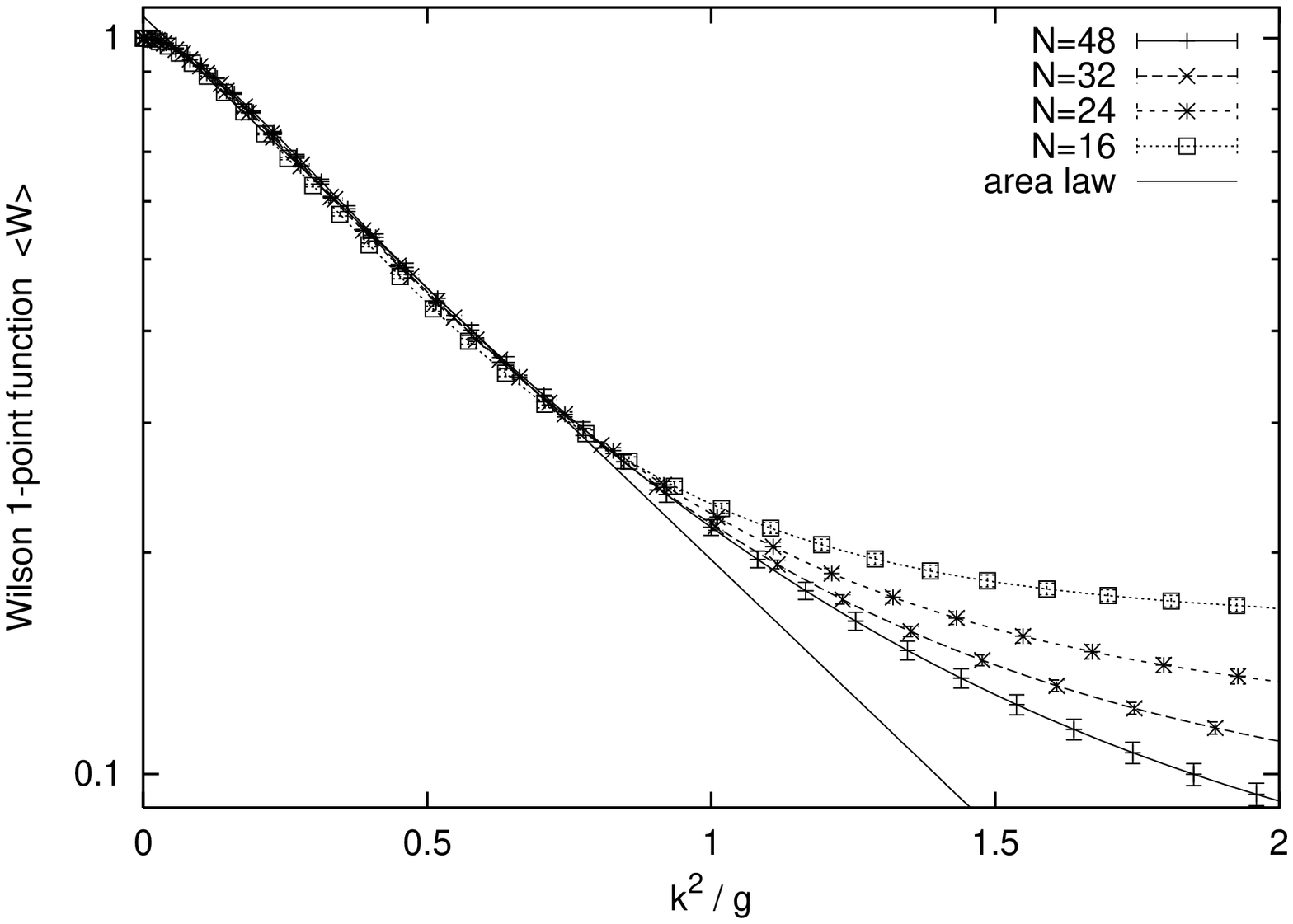}
    \caption{The Wilson 1-point function $\langle W \rangle$
             is plotted now logarithmically
             against $k^{2}/g$, in order to visualize the extent of 
             the area law behavior.
The scale parameter $g$ has been tuned as described in the text.}
\label{fig:Wilson1-log}
  \end{center}
\end{figure}

In the small $k$ regime it can be expanded as
\beqa
\langle W(k) \rangle &=& 1 + \frac{1}{2N} k^4 
\langle \tr ([X_1 , X_2] ^2) \rangle + O(k^{6}) \n
&=& 1- \frac{1}{4}k^4 \left( N-\frac{1}{N} \right)
+ O(k^{6}) ,
\label{small_k}
\eeqa
where we have used the exact result (\ref{trF2}).
Therefore, in order to make the small $k$ regime scale,
we have to take $g \propto 1/\sqrt{N}$, as we mentioned above.
In Fig.\ \ref{fig:Wilson1} we plot $\langle W(k) \rangle$ against
$k/\sqrt{g}$.
The small $k$ region scales as it should,
and the results agree with the analytical prediction (\ref{small_k}).
Moreover the scaling extends up to $k/\sqrt{g} = O(1)$.

If the model is equivalent to ordinary gauge theory ---
namely to 4D pure super Yang-Mills theory with four supercharges --- 
which is confining, then
the Wilson loop should exhibit an area law behavior.
In order to illustrate this behavior,
we show a logarithmic plot of $\langle W(k) \rangle$ versus the
area $k^{2}/g$ in Fig.\ \ref{fig:Wilson1-log}.
In this figure only, we fine-tune $g$ as a function of $N$ so that
the scaling in the intermediate regime of $k$ becomes even better.
We stay with the convention  $g(48)=1$ and use the optimal values
$g(32)=1.291$, $g(24)=1.563$,
$g(16)=1.929$, which is not far from $g \propto 1/\sqrt{N}$.
The small deviation can be understood as a manifestation
of finite $N$ effects. Fig.\ \ref{fig:Wilson1-log} shows
indeed a region of $k$ that corresponds to the area law
behavior $\langle W(k) \rangle \sim \exp (- \mbox{const.} k^2)$.
Surprisingly, the area law behavior is also observed 
in the bosonic model,
as Fig.\ \ref{fig:Wilson1-log-bos} shows,
which is quite contrary to what one might have expected \cite{HNT,Potsdam}.
In both cases, supersymmetric and bosonic,
it is not clear from the data 
whether the area law extends to $k=\infty$ in the 
large $N$ limit.
We will discuss the observed area law behavior
from a theoretical point of view later.

\begin{figure}[htbp]
  \begin{center}
    \includegraphics[height=12cm,angle=270]{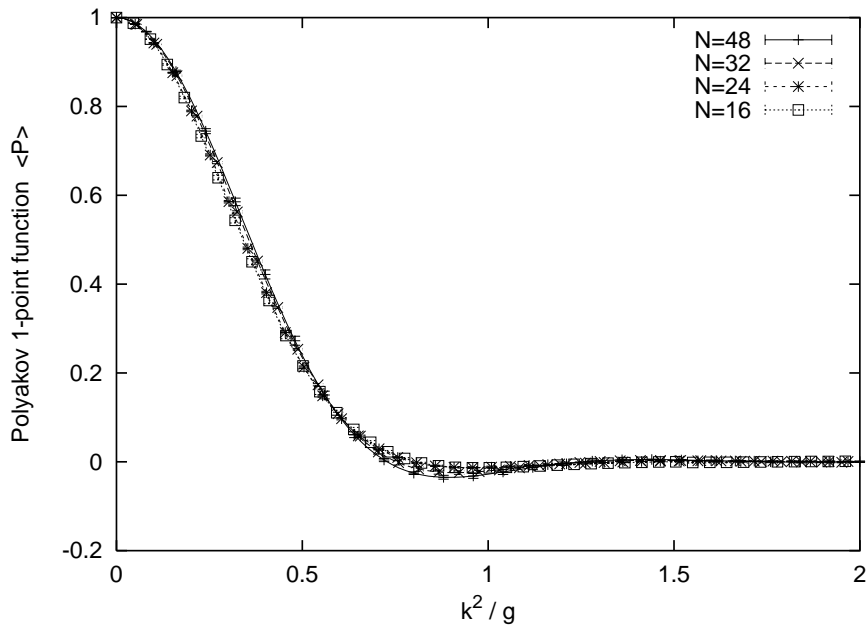}
 \caption{The Polyakov 1-point function $\langle P \rangle$, plotted
against $k/\protect\sqrt{g}$.}
\label{fig:Polyakov1}
  \end{center}
\end{figure}

We now proceed to the one-point function of the Polyakov line.
In the 2D Eguchi-Kawai model \cite{2DEK} this quantity
vanishes due to $(\IZ_N )^D$ symmetry.
In the present model, however, 
there is no exact symmetry that could make such a quantity vanish,
as we explained in Section \ref{model}.
Note for instance that $P(k=0) =1$ for any configuration.
Fig.\ \ref{fig:Polyakov1} shows the results for $\langle P(k)\rangle$.
It falls off rapidly as $k$ increases
\footnote{One may consider the small $k$ expansion here, as in eq.\ (\ref{small_k}).
The result is $\langle P(k) \rangle 
= 1 - \frac{1}{2N} k^2 \langle \tr (X_1 ^{~2})\rangle  + \cdots$.
The fact that $\langle 
\tr (A_\mu^{~2})\rangle$ is logarithmically
divergent means that actually $\langle P(k) \rangle$ has a
non-analytic behavior $\sim 1 + {\rm const.} \, k^2 \ln \vert k\vert $ around $k=0$.}.
Again we observe a good scaling with $g \propto 1/ \sqrt{N}$.
Remember also that $\langle P(k)\rangle$ is actually just
a Fourier transform of the eigenvalue distribution.
Therefore, the value of $k$ at which $\langle P(k)\rangle$ drops to zero,
which we denote as $k_0$,
should be inversely proportional to the space-time extent $R_{\new}$.
The observed scaling with $g \propto 1/ \sqrt{N}$ is consistent
with our result in the previous section,
$R_{\new}\sim \sqrt{g} \, N^{1/4}$.
The result for the bosonic case is shown in Fig.\ \ref{fig:Polyakov1-bos}.
We obtain a similar behavior except for some oscillations
in the large $k$ region. In particular, scaling is confirmed
with $g \propto 1/\sqrt{N}$.
The value of $k_0$ is larger than the supersymmetric $k_0$,
as expected.
The ratio of $k_0$ in the two cases is indeed roughly 
the inverse of the corresponding ratio of $R_{\new}$
(the bosonic $k_{0}$ is about twice as large as the 
supersymmetric one).

The above observations concerning $\langle P(k)\rangle$ and
$R_{\new}$ have an interesting implication 
on the Eguchi-Kawai equivalence.
We recall that from the results for $\langle W(k)\rangle$,
we phenomenologically concluded that 
the Eguchi-Kawai equivalence holds 
at least in a finite range of scale
for both, the supersymmetric and the bosonic case.
We would like to understand this from a theoretical point of view.
As we mentioned in Section \ref{model},
in the proof of Eguchi-Kawai equivalence,
$\langle P(k)\rangle$ is assumed to vanish.
We have found that $\langle P(k)\rangle$ is indeed
small for $k > k_0$, but not for $k < k_0$.
This means that the proof works for $k > k_0$,
but not for small $k$, which corresponds to
the ultraviolet regime in the corresponding gauge theory.
We also observed that $k_0$ remains finite with respect to
a physical scale in the large $N$ limit.
A complementary understanding can be obtained by
taking Gross-Kitazawa's point of view \cite{GK}.
As explained in Ref.\ \cite{HNT}, 
the extent of the eigenvalue distribution of $A_\mu$
determines the momentum cutoff of the corresponding
gauge theory \cite{GK}.
The observation in Section \ref{spacetime}
that $R_{\new} \sim \sqrt{g} \, N^{1/4}$
implies that the momentum cutoff remains finite
in physical scale as $N \to \infty$.
Let us assume
that the momentum cutoff is finite, but large enough to
attract the renormalization flow to
the fixed point which corresponds
to the universality class of gauge theory.
Then the flow follows closely the renormalization trajectory 
of the gauge theory, in a certain regime.
That would explain why the equivalence holds
at least in a finite range of scale.
However, since the momentum cutoff does not go to infinity in the
large $N$ limit, it is conceivable that
the renormalization flow will leave 
the renormalization trajectory of the gauge theory
at some low-energy scale eventually.
In this case the observed area law would not
extend to $k =\infty$ even in the large $N$ limit.

\subsection{Multi-point functions and universal scaling}
\label{correlators}

In this subsection we proceed to the large $N$ scaling
of multi-point functions of Wilson loops.
We first note that in the bosonic case, there are analytical results
to all order in the $1/D$ expansion \cite{HNT}. The statement is that
\beq
\langle {\cal O }_1{\cal O} _2 \cdots {\cal O} _n \rangle _{con}
\sim O\left(\frac{1}{N^{2(n-1)}}\right) ~~~~~\mbox{for the bosonic case},
\label{bosonicscaling}
\eeq
where ${\cal O}_i$ denotes a Wilson loop or a Polyakov line
as defined in eq.\ (\ref{wilsondef}), and
$\langle \cdots  \rangle _{con}$ means that only the connected part is taken.
The correlation functions should be considered as functions of 
$k_{\phys} = k/\sqrt{g}$, where
$g$ is taken to be proportional to $1/\sqrt{N}$.
Our results for the bosonic model shown in Figs. \ref{fig:Wilson1-bos} 
to \ref{fig:wil4-bos} clearly confirm this analytical prediction.
Let us consider a wave-function renormalization for each operator,
${\cal O}^{{\rm (ren)}}_{i} = Z {\cal O}_{i}$, 
so that connected correlation functions of the renormalized operators
${\cal O}^{{\rm (ren)}}_i$ become finite in the large $N$ limit.
Relation (\ref{bosonicscaling}) means, however, that
we cannot make all the multi-point functions finite.
If we make the two-point functions finite by choosing
$Z \sim O(N)$, then all the higher-point functions vanish 
in the large $N$ limit.
In the supersymmetric case, we will see that scaling is observed
again with $g \propto 1/\sqrt{N}$, but in contrast to the bosonic
case a universal $Z$ that makes all the correlators finite
seems to exist.
In the following, we always set $Z(N=48)=1$, without
loss of generality.

\begin{figure}[htbp]
  \begin{center}
    \includegraphics[height=12cm,angle=270]{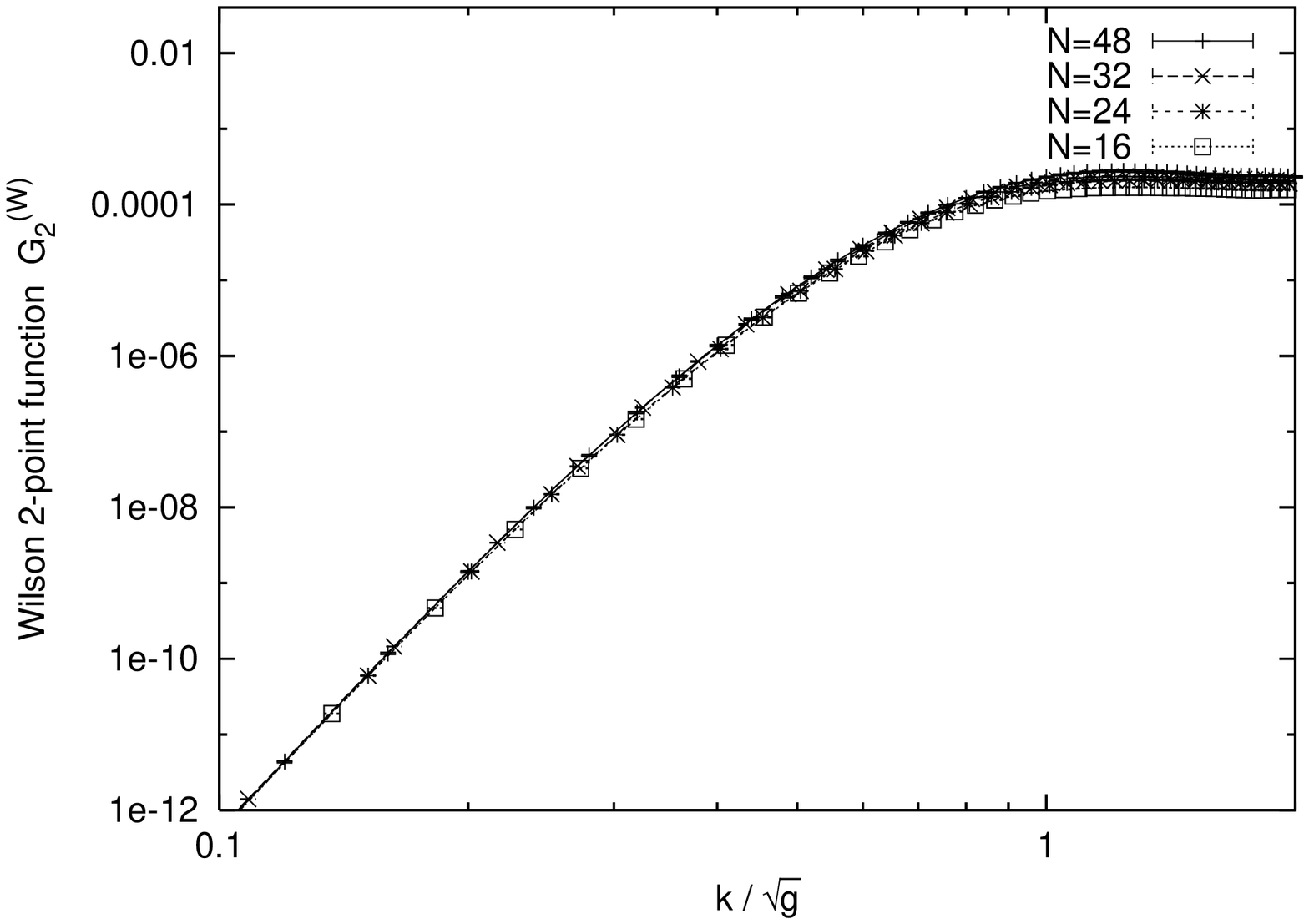}
 \caption{ The Wilson 2-point function $G_{2}^{(W)}$,
multiplied by $Z^{2} \propto N^2$, plotted against $k/\protect\sqrt{g}$.}
\label{fig:Wilson2-}
  \end{center}
\end{figure}

\begin{figure}[htbp]
  \begin{center}
    \includegraphics[height=12cm,angle=270]{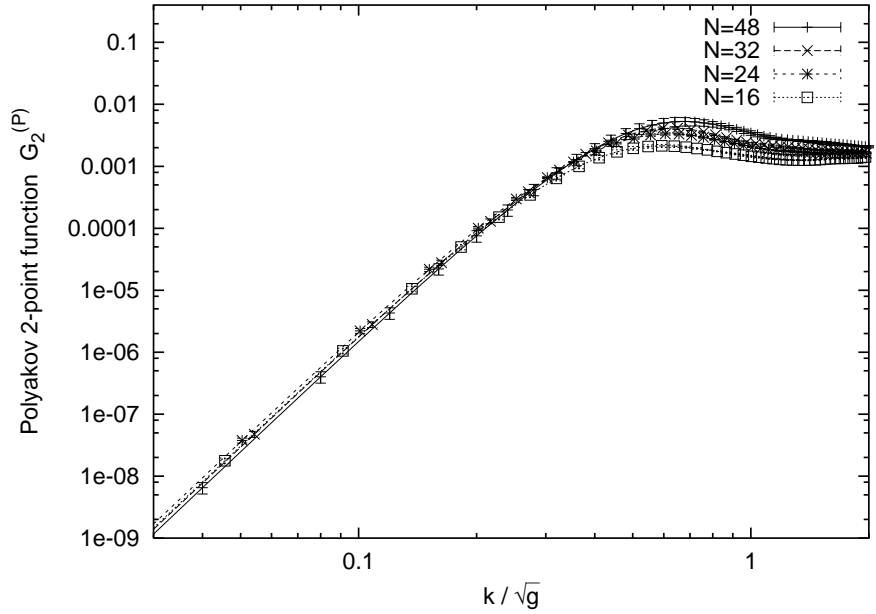}
 \caption{The Polyakov 2-point function $G_{2}^{(P)}$, 
multiplied by  $Z^{2} \propto N^2$, 
plotted against $k/\protect\sqrt{g}$.}
\label{fig:Polyakov2-}
  \end{center}
\end{figure}

Let us start with the two-point functions,
for which we measure the following two correlation functions,
\beqa
G_2^{(W)}(k) &=& \langle \{ \Imag  W(k) \} ^2  \rangle \n
G_2^{(P)}(k) &=& \langle \{ \Imag P(k) \} ^2  \rangle  \ .
\label{opn-2pt}
\eeqa
We take the imaginary part in order to avoid subtraction
of a disconnected part. 
\footnote{We also measured a number of multi-point functions, 
which are not presented here since the relative errors are rather large.}
(Note in this regard that since $\Imag  W(k)$ and $\Imag  P(k)$ 
are parity odd,
the one-point functions $\langle \Imag  W(k)\rangle$ 
and $\langle \Imag  P(k)\rangle$ vanish
due to parity invariance of the model.)
The results are shown in Figs.\ \ref{fig:Wilson2-}
and \ref{fig:Polyakov2-}, respectively.
If we multiply the data by $(N/48)^2$,
they scale nicely with $g \propto 1/\sqrt{N}$.

\begin{figure}[hbt]
   \begin{tabular}{cc}
      \hspace{-0.6cm}
\def\fpsangle{270} \epsfysize=85mm \fpsbox{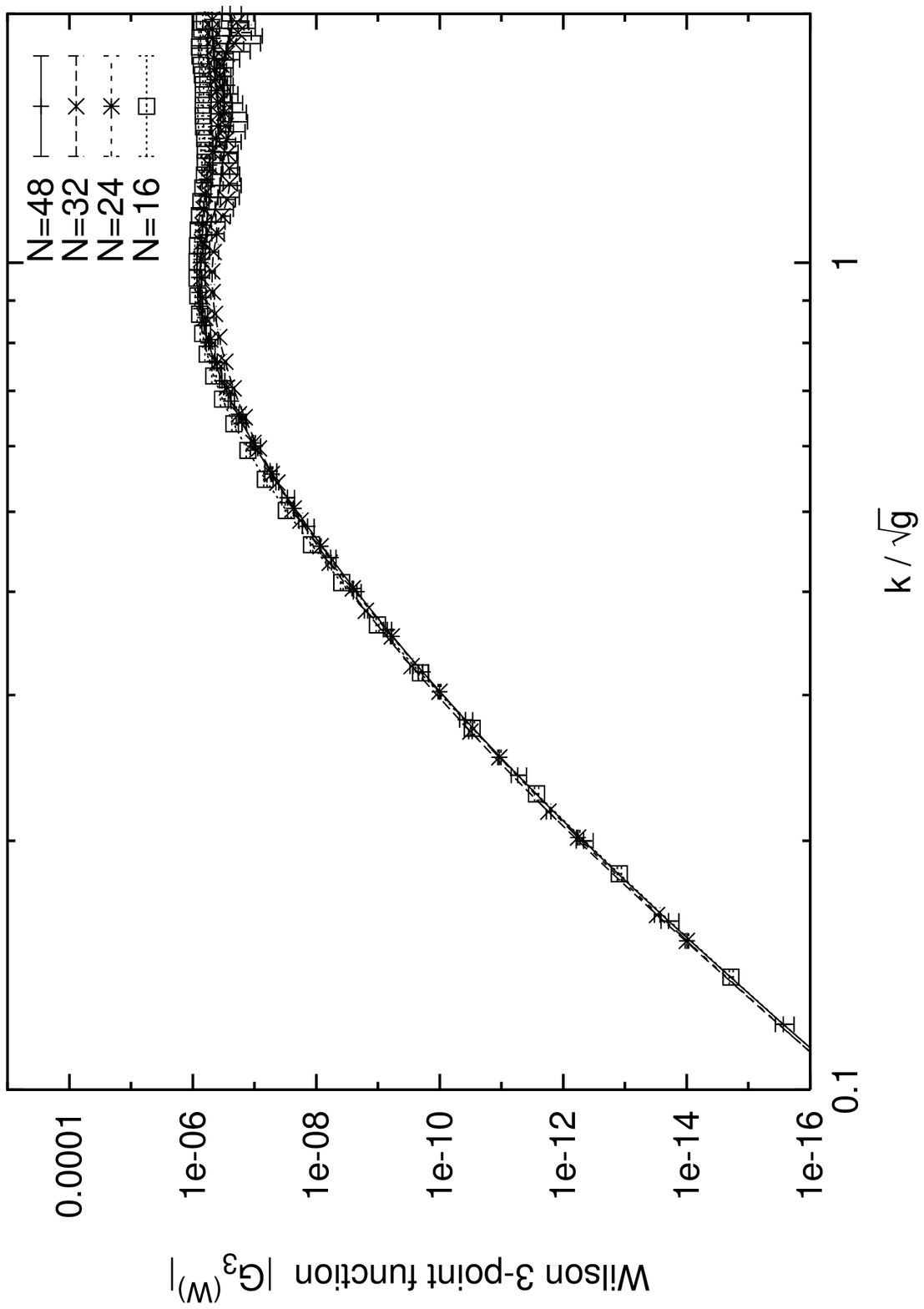} &
      \hspace{-0.6cm}
\def\fpsangle{270} \epsfysize=85mm \fpsbox{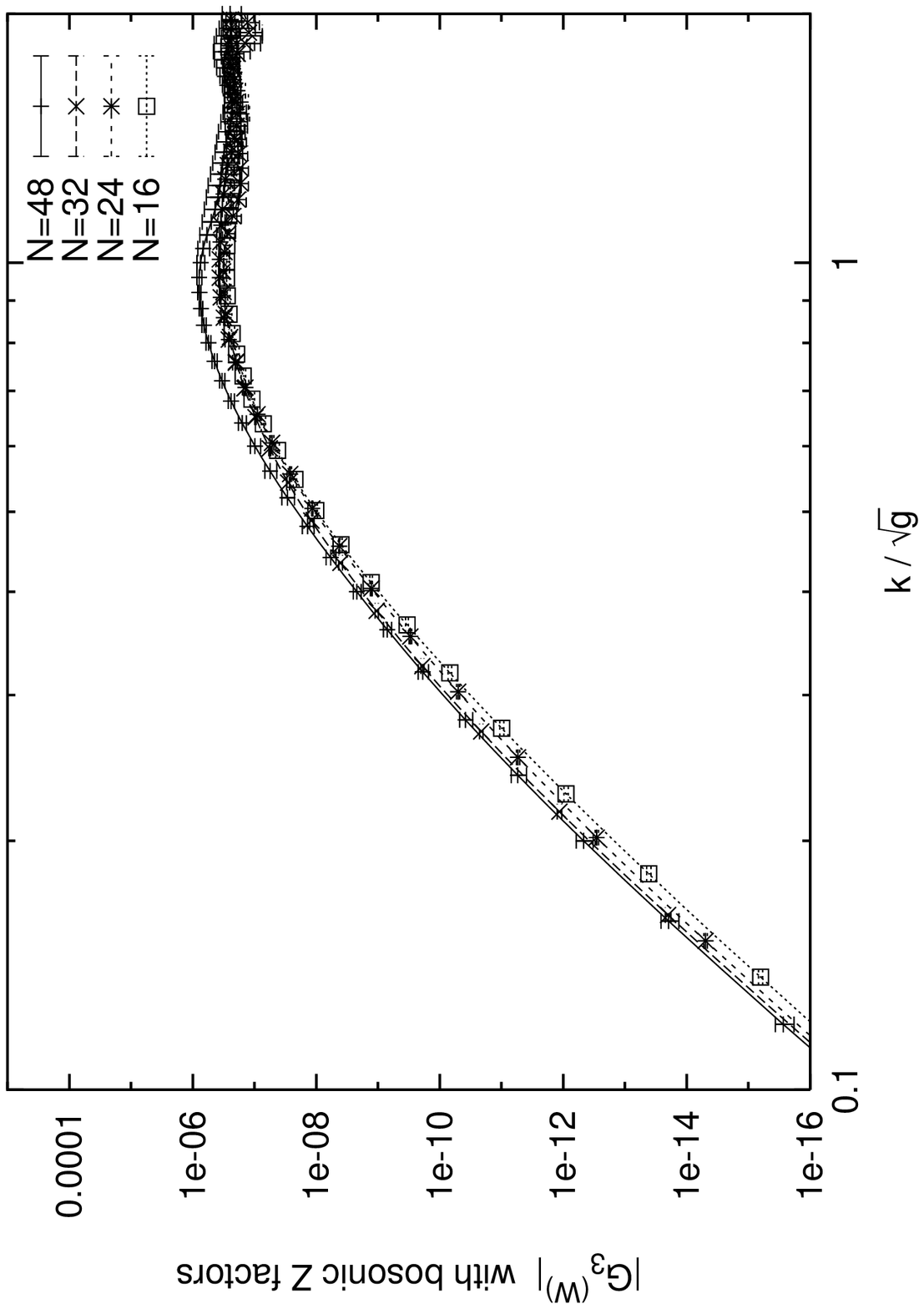}
   \end{tabular}
\caption{The Wilson 3-point function $G_{3}^{(W)}$,
multiplied by $Z^{3} \propto N^3$, plotted against $k/\protect\sqrt{g}$
on the left. On the right we show the corresponding plot using
the bosonic prediction $Z^{3} \propto N^4$ instead, which leads to
an inferior level of scaling.}
\label{fig:Wilson3}
\end{figure}

As a three-point function, we measure
\beq
  G_3^{(W)}(k) =
\langle (\Imag W(k))^2 \Real W(k) \rangle
- \langle (\Imag W(k))^2  \rangle 
 \langle \Real W (k)  \rangle   .
\eeq
We multiply the data either by 
$(N/48)^3$, which is required for the universal
scaling of all the multi-point correlation functions, or by
$(N/48)^4$, which is the factor predicted for the
bosonic model. The results are compared in Fig. \ref{fig:Wilson3}.
We do observe a nice scaling behavior with a factor of $(N/48)^3$,
but the scaling becomes worse for a factor of $(N/48)^4$.

\begin{figure}[hbt]
   \begin{tabular}{cc}
      \hspace{-0.6cm}
\def\fpsangle{270} \epsfysize=85mm \fpsbox{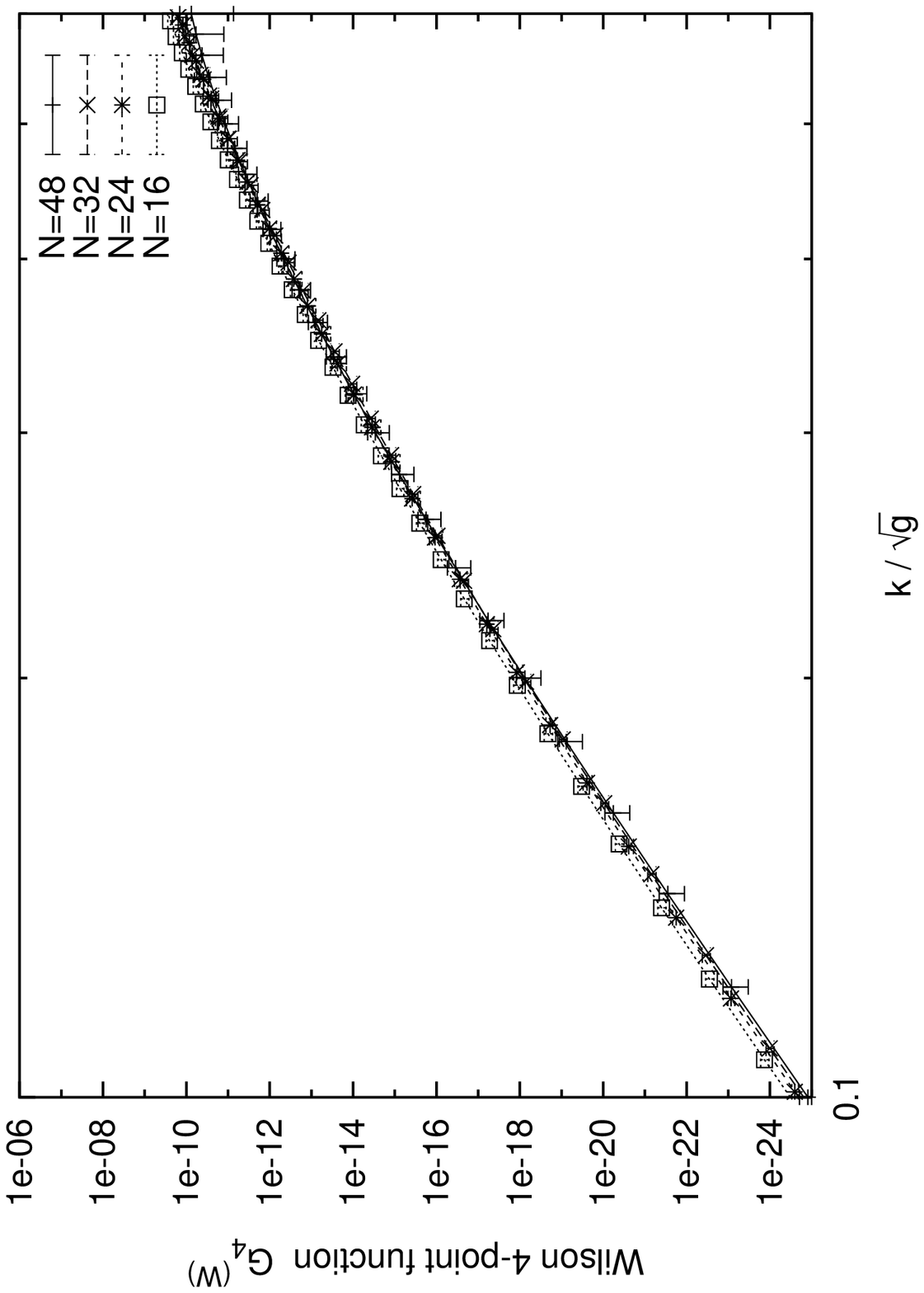} &
      \hspace{-0.6cm}
\def\fpsangle{270} \epsfysize=85mm \fpsbox{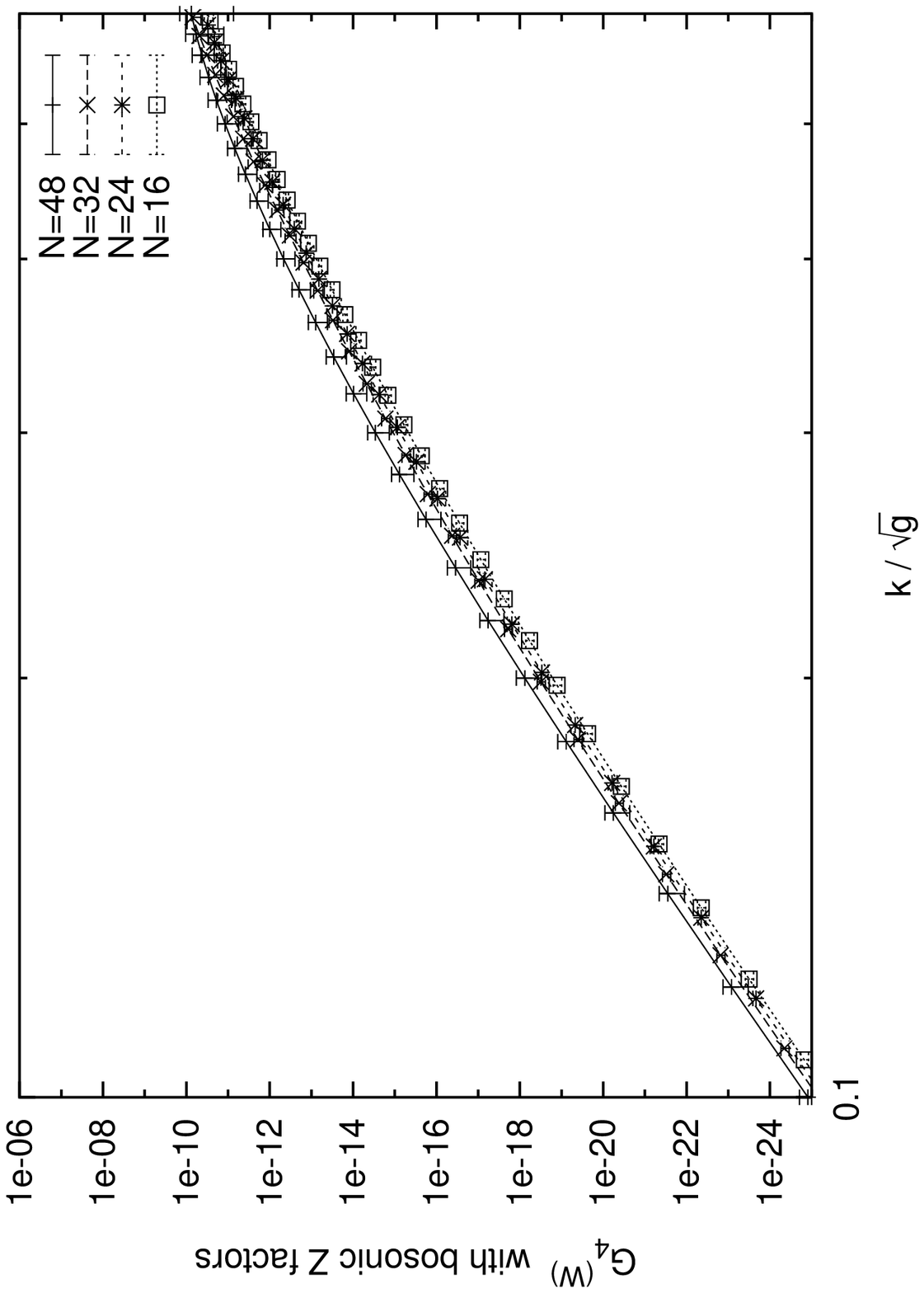}
   \end{tabular}
\caption{The Wilson 4-point function $G_{4}^{(W)}$,
multiplied by $Z^{4} \propto N^4$, plotted against $k/\protect\sqrt{g}$
on the left. On the right we show the corresponding plot using
the bosonic prediction $Z^{4} \propto N^6$ instead, which leads to
an inferior level of scaling.}
\label{fig:Wilson4}
\end{figure}

Similarly, as a four-point function we measure
\beq
  G_4^{(W)}(k) =
\langle (\Imag W(k))^4 \rangle
- 3 \langle (\Imag W(k))^2  \rangle .
\eeq
We multiply the data either by $(N/48)^4$, which is required for the universal
scaling of all the multi-point correlation functions, or by 
$(N/48)^6$, which is the factor predicted for the bosonic model.
The results are compared in Fig. \ref{fig:Wilson4}.
Again the scaling behavior obtained with the factor for universal scaling
is superior over the behavior with the bosonic factor.

To summarize our results concerning Wilson loop correlators, 
we observe that
\beqa
&~&\langle {\cal O }\rangle  \sim O(1) \n
&~& 
 \langle {\cal O }_1{\cal O} _2 \cdots {\cal O} _n \rangle _{con}
\sim O\left(\frac{1}{N ^n}\right) ~~~~~\mbox{for}~~n\ge 2.
\label{summarymulti}
\eeqa
These correlators scale as functions of 
$k_{\phys} = k/\sqrt{g}$, where
$g$ is taken to be proportional to $1/\sqrt{N}$.
This means that all the multi-point functions
of the renormalized operators ${\cal O}^{{\rm (ren)}}_{i} = Z {\cal O}_{i}$
become finite in the large $N$ limit if we set $Z \sim O(N)$,
in contrast to the bosonic case.
We will discuss further the implications of this universal scaling
behavior in the next subsection.

Finally we comment on large the $N$ factorization.
In ordinary gauge theory, large $N$ factorization
can be shown by weak-coupling expansion as well as
strong-coupling expansion.
In a large $N$ reduced model with Hermitian matrices, 
one cannot do a weak-coupling or a strong-coupling expansion,
because $g$ is not a coupling constant but a scale parameter,
as we have mentioned. Hence large $N$ factorization is 
nontrivial.
In the bosonic case, large $N$ factorization holds to all 
orders of the $1/D$ expansion \cite{HNT}.
Our observation (\ref{summarymulti}) implies
\beq
\langle {\cal O }_1{\cal O} _2 \cdots {\cal O} _n \rangle
=  \langle {\cal O} _1\rangle\langle
{\cal O }_2 \rangle \cdots \langle{\cal O }_n \rangle
+ O \left( \frac{1}{N^{2}}\right) \ ,
\eeq
where the $O \left( 1/N^{2}\right)$ contributions
are due to $ \langle {\cal O} _1 {\cal O }_2 \rangle _{con} 
 \langle {\cal O} _3\rangle \cdots \langle{\cal O }_n \rangle$, etc.
Therefore the large $N$ factorization is also valid 
in the supersymmetric case.

\subsection{Interpretation of the large $N$ scaling}
\label{interpretation}

In this subsection, we further discuss the physical significance of the 
large $N$ scaling (\ref{summarymulti}) we observed.


If one views the supersymmetric matrix model considered here 
as a non-perturbative definition of type IIB string theory, and 
the Wilson loops as fundamental strings, string 
unitarity requires a large $N$ behavior of the form $N^{a\chi _n}$
for the connected correlators of $n$ Wilson loops,
where $\chi_n= 2-n$ is the Euler characteristic
of the worldsheet. 
In order to compare our results
for the supersymmetric case (\ref{summarymulti}) as well as 
for the bosonic case (\ref{bosonicscaling})
to this behavior, 
we first drop the extra $1/N^n$ factor in (\ref{bosonicscaling})
and (\ref{summarymulti}),
which is due to the chosen normalization (\ref{wilsondef})
of the operators ${\cal O}_i$. 
Then we find that the connected correlators of 
Wilson loops change from an $O(N^{\chi_n})$ behavior 
to an $O(1)$ behavior by the introduction of supersymmetry.
Our results for the supersymmetric case indicate $a=0$, 
which is not in contradiction to the above requirement of
string unitarity, but it is an extreme case where one 
is far away from a perturbative expansion in genus. It would 
be interesting to understand the result $a=0$ analytically, directly 
from the matrix model. From the string theoretical point of view
this indicates that the supersymmetric matrix model might automatically
realize a kind of double scaling limit, as it has been 
known from (ordinary) matrix models of two-dimensional quantum gravity.
In these models contributions from all genera are important.

While we do not presently have an analytic understanding of the difference 
between the connected correlators in
the supersymmetric and the bosonic case, we can 
try to make an educated guess, based on the perturbative 
expansion (\ref{decompose}).  

Let us consider the bosonic case.
After integrating out the off-diagonal elements perturbatively,
schematically each diagram gives a contribution
\beq
\sum _ {i_1,\cdots ,i_F} 
\left(  \frac{g^2}{(x_i - x_j)^2}  \right) ^L
\left(  (x_k - x_l ) \frac{1}{g^2 }  \right) ^{V_3}
\left(  \frac{1}{g^2}  \right) ^{V_4} \ ,
\label{diagram_before}
\eeq
where $x_i$ denotes the diagonal elements as usual, and
$F$, $L$, $V_3$, $V_4$ are 
the number of index loops (faces), propagators (links), 
3-point and 4-point vertices, respectively.
They obey the relations
\beq
F+V-L = \chi \n \ , \qquad
4 V_4 + 3 V_3 = 2 L \n \ , \qquad
V = V_3 + V_4 \ .
\eeq
Here $\chi$ is the Euler characteristic of the diagram given 
as $\chi = 2 - 2 h - n$, where $h$ is the genus (the number of 
handles in the diagram) and $n$ is the number of the operators.

Let us now integrate over the diagonal elements $x_i$,
under the assumption that the infrared region dominates.
This assumption implies that the $x_i$'s are not close. In fact 
we assume that they are generically separated by some scale $R$, 
which is a measure of the extension of our universe. Viewing the 
$x_i$'s as the coordinates of the worldsheet in target 
space, the above assumption implies that the worldsheet is 
{\it rough}: points are scattered quite randomly.
We can now estimate the value for the diagram considered
by simply replacing the integration over the $x_i$'s by their 
typical separation, assuming the existence of a suitable measure 
which implements the above hypothesis. We obtain
\beq
N ^F 
\left(  \frac{g^2}{R^2}  \right) ^L
\left(  \frac{R}{g^2 }  \right) ^{V_3}
\left(  \frac{1}{g^2}  \right) ^{V_4}  \sim N ^\chi \ ,
\eeq
where we have used $R \sim \sqrt{g} \, N^{1/4}$ .
So the assumption of infrared dominance
reproduces the observed large $N$ behavior. 

In the supersymmetric case,
the fermion diagonal elements make the power-counting 
more complicated. 
However, let us naively consider
the contribution (\ref{diagram_before}).
When we integrate over the diagonal elements $x_i$,
let us assume that the {\em ultraviolet} rather than the infrared region 
dominates. Namely, we assume that
{\it smooth} worldsheets are favoured dynamically due to 
supersymmetry 
\footnote{It should be remarked here that such an 
increased smoothness of the bosonic part in a supersymmetric 
worldsheet has been observed in toy models for superstrings \cite{av}.}.
More precisely we assume that the dominant $x_i$ configurations are 
such that $x_i$'s, which are connected in a given diagram, are as close 
together as the actual distribution of $x_i$'s allows. Above we have 
seen that there seems to be such a {\it minimal length} \ $\ell$,
which is of the order $\sqrt{g}$,
characterizing the distribution of $|x_i-x_j|$. Replacing $|x_i-x_j|$
with this minimal length in (\ref{diagram_before}) we arrive at
\beq
\left(  \frac{g^2}{\ell^2}  \right) ^L
\left(  \frac{\ell}{g^2 }  \right) ^{V_3}
\left(  \frac{1}{g^2}  \right) ^{V_4}  \sim  O(1) \ .
\eeq
Thus the assumption of ultraviolet dominance
leads to our observed results in the supersymmetric case.
Of course it remains to be understood why there is a difference 
between infrared and ultraviolet dominance in the bosonic and 
supersymmetric cases.

If the above naive argument is true, it also implies that the contributions
in the supersymmetric case do not depend on the genus $h$ either, 
and consequently that diagrams of all  topologies
contribute with equal weight.
This is reminiscent of the double scaling limit of matrix models.
For example, let us consider a Hermitian one-matrix model with the
partition function
\beq
Z = \int \dd \phi \ \ee ^{- N (\trs \phi ^2 - \lambda \trs \phi ^3)} ,
\eeq
$\phi$ being a $N \times N$ Hermitian matrix.
The $n$-point correlation functions of loop operators behave as
\beq
\langle \tr \phi ^{l _1} \cdots \tr \phi ^{l _n} \rangle
\sim (N \epsilon ^{5/2})^{2-2h -n} \epsilon^{-n} \ ,
\eeq
where $\epsilon$ is the parameter related to the $l_i$ and to
the coupling constant $\lambda$ as
\beq
l_i \sim \frac{1}{\epsilon} ~~~~~ ,
~~~~~\lambda - \lambda _c  \sim \epsilon^2  \ ,
\eeq
and $\lambda _c$ is the critical coupling constant.
Note that the large $N$ behavior for fixed $\epsilon$ is $N ^{\chi}$,
which we obtain for the bosonic large $N$ reduced model.
When one takes the large $N$ limit with 
$N \epsilon ^{5/2}=g_{{\rm str}}^{-1}$ fixed,
the dependence on $h$ disappears and the remaining power behavior
$\epsilon ^{-n}$ can be absorbed into the wavefunction renormalization of 
each operator.
In the large $N$ reduced model,
we do not have the parameter corresponding to $\lambda$.
The large $N$ scaling we observed 
for the supersymmetric case
is formally the same as one obtains 
in the double scaling limit of the one-matrix model.
In this sense, one might say that 
the double scaling limit is taken automatically in the 
supersymmetric large $N$ reduced model,
and that the string coupling constant $g_{{\rm str}}$ is not a tunable
parameter but is fixed dynamically.
If this is the case, it should be considered as a very satisfactory
feature of the supersymmetric large $N$ reduced model
as a nonperturbative definition of superstring theory,
since we do expect
the string coupling constant $g_{{\rm str}}$, which is related to 
the vacuum expectation value of the dilaton field,
to be fixed dynamically (if superstring theory is treated
nonperturbatively).
It is also interesting that the qualitative difference of the 
large $N$ behavior for the bosonic and the supersymmetric case
might be traced back to the smoothness of the worldsheet,
which is itself a dynamical question to be addressed.
This point needs further clarification.

\section{Summary and discussion}
\setcounter{equation}{0}
\label{summary}

In this paper, we have studied the
large $N$ dynamics of a supersymmetric large $N$ reduced model
by means of Monte Carlo simulations.

We studied the space-time structure represented by the eigenvalues
of the bosonic matrices.
In particular, we found that the large $N$ power behavior of 
the space-time extent is consistent with the branched-polymer 
picture based on the one-loop perturbative expansion
around diagonal matrices.
The effect of fermions in the space-time extent
was observed by the enhancement of 
the coefficient in the power behavior, but not in the power itself.
The power appears to be the same for the bosonic and supersymmetric case.
We emphasized, however, that 
the theoretical explanation is completely different.
We also found that the space-time uncertainty 
is clearly reduced for the supersymmtric case, which means that
space-time comes closer to the classical behavior.
Even in the supersymmetric case,
the space-time uncertainty is found to be finite 
in the physical scale in the large $N$ limit.
We argued that this implies that the model satisfies the uncertainty
principle for the nonperturbative definition of superstring theory.

The large $N$ scaling behavior of Wilson loop correlators
is observed at fixed $g^2 N$.
Although this scaling of $g$ is the same as
in the bosonic model, there is
a striking difference from the bosonic case
in the wave-function renormalization
with the multi-point functions.
In the bosonic case, there was no universal scaling behavior:
keeping two-point functions finite,
all the higher-point functions vanish.
In the supersymmetric case, we observed a clear trend
for all the higher-point functions to become finite in the
large $N$ limit.
We gave a perturbative argument that 
this result for the supersymmetric case 
might be understood if we assume smooth worldsheets to dominate.
This argument also implies that all the topologies of the worldsheet
contribute with equal weight to the amplitude.
All these features are reminiscent of the double scaling limit
of matrix models. 

We also addressed the issue of Eguchi-Kawai equivalence.
By searching for the area law behavior in the 
one-point function of the Wilson loop,
we concluded that the equivalence does hold
at least in a finite region of scale.
What is rather surprising is that the area law behavior
has been observed also for the bosonic model.
This suggests that the bosonic model is also equivalent
to ordinary large $N$ Yang-Mills theory 
at least in a finite region of scale,
which is contrary to what has been generally believed.
We argued, however, that this conclusion can be understood 
from a more theoretical point of view 
based on the large $N$ behavior obtained for $R_{\new}$ and 
the one-point function of the Polyakov line.
It is an open question whether this equivalence extends to the far
infrared regime.

To summarize, we have gained new insight into the dynamical properties
of the large $N$ behavior of a supersymmetric large $N$ reduced model.
We hope that our findings shed light on the dynamical aspects
of the most interesting 10D version of our model,
i.e.\ the IIB matrix model.
In this respect, it is encouraging that
the large $N$ scaling of Wilson loop correlators
in the present model has been observed at fixed $g^2 N$, 
which coincides with the result obtained by requiring that 
the loop equations of the IIB matrix model should reproduce 
the string field Hamiltonian.
We presume that a large $N$ scaling of Wilson loop correlators
--- like the one we observed --- also holds for the IIB matrix model;
then the only difference would be the spontaneous breakdown
of Lorentz symmetry.
One of the good news revealed in the present work
is that low energy effective theory,
based on the one-loop approximation,
does already capture the low energy dynamics 
of the supersymmetric matrix model.
We therefore hope to address the most
interesting issue of spontaneous breakdown of Lorentz invariance
by using the low energy effective theory --- which is in 10D 
far more complicated than in the 4D case.
We are going to report on Monte Carlo studies of IIB matrix model
along these lines in the near future.


\section*{Acknowledgment}
We thank T.\ Ishikawa, C.F.\ Kristjansen, Y.M. Makeenko, T.\ Nakajima,
M. Staudacher, A. Tsuchiya and G. Vernizzi for valuable discussions.
J.\ N.\ is supported by the Japan Society for the Promotion of
Science as a Postdoctoral Fellow for Research Abroad.  
The computation has been done partly on 
Fujitsu VPP500 at High Energy Accelerator Research Organization (KEK),
Fujitsu VPP700E at The Institute of Physical and Chemical Research (RIKEN),
and NEC SX4 at Research Center for Nuclear Physics (RCNP) of Osaka
University.
This work is supported by the Supercomputer Project (No.99-53)
of KEK.


\vspace*{1cm}


\appendix

\section{The algorithm for the Monte Carlo simulation}
\setcounter{equation}{0}
\label{algorithm}
\renewcommand{\theequation}{A.\arabic{equation}}
\hspace*{\parindent}
In this appendix, we explain the algorithm we use
for the Monte Carlo simulation of the supersymmetric matrix model.
Only in this appendix we set $g=1$ for simplicity.

We first carry out the integration over fermionic matrices
to obtain the explicit formula for the fermion determinant.
We calculate
\beq
Z_f [A] = \int \dd \psi \dd \bar{\psi}  ~ \ee ^{-S_f } ,
\label{Zfdef}
\eeq
where we use the notation introduced in eq.\ (\ref{action}).
We define a set of generators
$t^a \in $ gl($N$,$\IC$) by
\beq
(t^a)_{ij} = \delta_{i i_a} \delta_{j j_a} 
~~~~~(a=1,\cdots,N^2),
\eeq
where $i_a$ and $j_a$ are integers running from 1 to $N$,
specified uniquely by
\beq
a = N (i_a -1) + j_a .
\label{map}
\eeq
We also introduce the notation
$\bar{a} = N (j_a -1) + i_a$.
The fermionic matrix $\psi _\alpha$
can be 
expanded in terms of $t^a$ as
\beq
(\psi_\alpha ) _{ij} = \sum_{a=1}^{N^2} 
\psi_{a \alpha} ~ (t^{a})_{ij} , 
\label{texpand}
\eeq
where $\psi_{a \alpha} = (\psi_{\alpha})_{i_a j_a}$.
$\bar{\psi}_{\alpha}$ and $A_{\mu}$ can be expanded similarly with
the coefficients $\bar{\psi}_{a \alpha}  = (\bar{\psi}_{\alpha})_{i_a j_a}$
and $A_{a \mu} = (A_{\mu}) _{i_a j_a} $.
Note also that $A_{\bar{a} \mu} = (A_{a \mu} )^* $ due to
the Hermiticity of $A_\mu$.

We define the structure constants $g_{abc}$ of \ gl($N$,$\IC$) by
\beqa
g_{abc} &=& \tr (t^c [t^a , t^b])  \n
&=& \delta_{j_a i_b} \delta_{j_b i_c} \delta_{j_c i_a} -
\delta_{j_c i_b} \delta_{j_b i_a} \delta_{j_a i_c} .
\eeqa
The fermionic action then reads
\beqa
S_f &=&  - g_{abc} \bar{\psi} _{c \alpha} 
(\Gamma_\mu)_{\alpha \beta}
A_{a \mu} \psi _{b \beta}  \n
&=&  - \bar{\psi} _{a \alpha} 
{\cal M }'_{a \alpha ,  b \beta} \psi _{b \beta} ,
\eeqa
where
\beq
{\cal M}'_{a \alpha ,  b \beta} = 
- g_{abc} (\Gamma_\mu)_{\alpha \beta} A_{c \mu}  .
\label{Mprimedef}
\eeq
We first integrate out $(\psi_{\alpha})_{NN}$ and 
$(\bar{\psi} _{\alpha})_{NN}$ using the $\delta$ functions
in the measure (\ref{measure_fermion}).
We get a factor of $1/N^4$ followed by the replacements
\beq
(\psi_{\alpha})_{NN} \Rightarrow - \sum _{j=1} ^{N-1} (\psi_{\alpha})_{jj} \ ;
~~~
(\bar{\psi}_{\alpha})_{NN} \Rightarrow - \sum _{j=1} ^{N-1} (\bar{\psi}_{\alpha})_{jj}  
\eeq
in the fermionic action.
The integration over the remaining Grassmann variables
yields \ $\det {\cal M} $, where
${\cal M}$ is the $\, 2(N^2-1)$ $\times$ $2(N^2-1)\,$ complex matrix
defined by
\beq
{\cal M }_{a \alpha , b\beta} = {\cal M}'_{a \alpha , b \beta} 
- {\cal M}'_{N^2 \alpha , b\beta}  \delta _{i_a j_a} 
-{\cal M}'_{a \alpha , N^2 \beta} \delta _{i_b j_b}
\label{defMmat}
\eeq
(the indices $a$ and $b$ run from 1 to $N^2-1$). Thus, we obtain 
\beq
Z_f [A] = \frac{1}{N^4} \det {\cal M} \ .
\eeq

We first want to show that
the determinant \ $\det {\cal M}$ \ is
real positive \footnote{This has been already reported in Ref.\ \cite{KNS}
as a numerical observation. For related work, see Ref.\ \cite{Hsu}.}.
For this purpose we note that the matrix ${\cal M}$ satisfies
the identity $\sigma _2 {\cal M} \sigma _2 = {\cal M}^*$.
Hence if $\varphi _{a \alpha}$ is an eigenvector of 
${\cal M}$ with an eigenvalue $\lambda$,
then $\psi _{a \alpha} = (\sigma _2)_{\alpha\beta} (\varphi _{a \beta})^*$ 
is an eigenvector of ${\cal M}$ with an eigenvalue $\lambda ^*$.
It is important that the two vectors $\varphi _{a \alpha}$,
$\psi _{a \alpha}$ are linearly independent.
The determinant, which is the product of all the eigenvalues of
${\cal M}$, should therefore be real and positive semi-definite.
In the case of 6D or 10D (IIB matrix model) versions of the supersymmetric
large $N$ reduced model, the fermion integral yields a complex
effective action in general.
This causes the notorious sign problem, which makes 
standard Monte Carlo simulations practically inapplicable for large $N$.
In the present case, since the determinant $\det {\cal M}$ is
real positive,
we can introduce a $\, 2(N^2-1)$ $\times$ $2(N^2-1) \,$
Hermitian matrix ${\cal D}= {\cal M}^\dag {\cal M}$,
which has real positive eigenvalues,
and $\det {\cal M} = \sqrt{\det {\cal D}}$. Therefore we have written
the effective action for the bosonic matrices $A_\mu$
in eq.\ (\ref{eff_act}) as
\beq
S_{\eff} = S_b -  \frac{1}{2} \ln \det {\cal D} \ .
\eeq
We apply the Hybrid R algorithm \cite{hybridR} to simulate this
system
\footnote{Ref.\ \cite{Weingarten} gives an overview of 
effective algorithms for dynamical fermions, including the Hybrid R
algorithm.}.

The first step of the Hybrid R algorithm is
to apply the molecular dynamics method \cite{molecule}. 
We introduce a conjugate momentum for $A_{a \mu}$ as
$X_{a\mu}\,$, which satisfies 
$X_{\bar{a}\mu} = (X_{a\mu} )^* $.
The partition function can be re-written as
\beq
Z = \int \dd X  \dd A ~ \ee ^{-H} ,
\eeq
where $H$ is the ``Hamiltonian'' defined by
\beq
H = \frac{1}{2} \sum _{\mu a} X_{\bar{a}\mu} X_{a \mu}
+ S_b [A] -   \frac{1}{2} \ln \det {\cal D}  .
\eeq
The update of $X_{a\mu}$ can be done by just
generating $X_{a\mu}$ with the probability distribution
$\exp (-\frac{1}{2}\sum |X_{a\mu}|^2 )$.
In order to update $A_{a \mu}$, we use the Hamiltonian equations
\beqa
\frac{\dd A_{a \mu} (\tau)}{\dd \tau}
&=&  \frac{\del H}{\del X_{a\mu}  }  = X_{\bar{a}\mu} ,  \\
\frac{\dd X_{a \mu} (\tau)}{\dd \tau}
&=& - \frac{\del H}{\del A_{a \mu}  }  
= \frac{1}{2} \tr \left(
 \frac{\del {\cal D} }{\del A_{a\mu} } {\cal D}^{-1}  \right)
- \frac{\del S_b}{\del A_{a \mu} }   .
\label{Hamiltonianeq}
\eeqa
Along the ``classical trajectory'' given by 
the Hamiltonian equation,

(i) $H$ is invariant, 

(ii) the  motion is reversible,

(iii) the phase-volume is preserved,
\beq
\frac{\del (A(\tau),X(\tau))}{
\del (A(0),X(0))} = 1 \ ,
\eeq
where $ (A(\tau),X(\tau))$ is a point on the 
trajectory after evolution from $(A(0),X(0))$.
Therefore, generating a new set of $(A,X)$ 
by solving the Hamiltonian equation for a fixed ``time'' interval 
$\tau$ satisfies detailed balance.
This procedure --- together with the proceeding
generation of $X_{a \mu}$  with the Gaussian distribution ---
is called ``one trajectory'', which corresponds
to ``one sweep'' in ordinary Monte Carlo simulations.

In order to solve the Hamiltonian equation numerically,
we have to discretize the ``time'' $\tau$.
A discretization which maintains
the properties (ii) and (iii) is known.
The slight violation of (i) for finite $\Delta \tau$
causes systematic errors.
One can in principle eliminate the systematic error completely, 
by making a Metropolis accept/reject decision at the end of each trajectory.
But in the present case, the overhead for this procedure
is rather large.
We therefore decided to omit that step, and just use
a sufficiently small $\Delta \tau$.
Still we can use the specific discretization of Ref.\ \cite{hybridR},
which we explain below, to minimize the systematic error.
As we explain later, we do find a good convergence 
in small $\Delta \tau$, and the systematic error is well under control.

We introduce a short-hand notation for the discretized
$X_{a\mu} (\tau)$ and $A_{a \mu} (\tau)$,
\beq
X_{a\mu} ^{(r)} = X_{a\mu}  (r \Delta \tau) ~~~;~~~~~
A_{a\mu} ^{(s)} = A_{a \mu} ( s \Delta \tau ) \ .
\eeq
The Hamiltonian equations are discretized as
\beqa
A_{a \mu} ^{(\frac{1}{4})} &=& A_{a\mu} ^{(0)}  + \frac{\Delta \tau }{4}
X_{\bar{a} \mu} ^{(0)} \nonumber \\
A_{a\mu} ^{(n+\frac{1}{2})} &=& A_{a\mu} ^{(n+\frac{1}{4})} 
 + \frac{\Delta \tau }{4} X_{\bar{a} \mu} ^{(n)} \nonumber \\
A_{a\mu} ^{(m+\frac{1}{4})} &=& A_{a\mu} ^{(m-\frac{1}{2})}
 + \frac{3 \Delta \tau }{4} X_{\bar{a} \mu} ^{(m)} \nonumber \\
A_{a\mu} ^{(\nu)} &=& A_{a\mu} ^{(\nu-\frac{1}{2})} + \frac{\Delta \tau }{2}
X_{\bar{a} \mu} ^{(\nu)} \nonumber \\
X_{a\mu} ^{(n+1)} &=& X_{a\mu} ^{(n)} 
+ \Delta \tau \left\{
\frac{1}{2} R_{a\mu} ^{(n+\frac{1}{2})}
-  \frac{\del S_b}{\del A_{a\mu} } 
\left(A_{a \mu} ^{(n+\frac{1}{2})} \right)
\right\} \quad ,
\label{traceDdD}
\eeqa
where $n=0,1,\cdots, \nu - 1$, $m=1,\cdots, \nu - 1$,
and $R_{a\mu} ^{(n+\frac{1}{2})}$ is defined by
\beqa  \label{Rdef}
 R_{c\mu} ^{(n+\frac{1}{2})}  
&=& \Phi _{a\alpha} ^* \left(  \frac{\del {\cal D} 
( A_{a\mu} ^{(n+\frac{1}{2})} )  }{\del A_{c\mu}} \right)_{a\alpha b \beta} 
\Phi _{b\beta} \ , \\
\label{invD}
{\cal D} ( A_{a\mu} ^{(n+\frac{1}{2})} )   \Phi &=& 
{\cal M}^\dag ( A_{a\mu} ^{(n+\frac{1}{4})} )   \eta  \ .
\eeqa
Here $\eta_{a\alpha} $ are complex variables
generated by the Gaussian distribution
$\exp ( - \sum _{a\alpha} |\eta_{a\alpha}|^2 )$.
The judicious choice of the argument of ${\cal M}^\dag$
is the tool to reduce the systematic error \cite{hybridR}.

We solve eq.\ (\ref{invD}) with respect to $\Phi$
by means of the conjugate gradient method \cite{CG},
which is iterative.
Each iteration involves a multiplication of the matrix ${\cal D}$
with some vector $v$.
Since ${\cal D}$ is a $\, 2(N^2 -1)$ $\times$ $2(N^2 -1)\, $ matrix,
storing ${\cal D}$ requires $O(N^4)$ memory,
and multiplying ${\cal D}$ with $v$ naively
involves $O(N^4)$ arithmetic operations.
Actually we can do much better than this.
We first recall that ${\cal D}= {\cal M}^\dag {\cal M}$,
where ${\cal M}$ is the $\, 2(N^2-1)$ $\times$ $2(N^2-1)\,$ matrix
defined in eq.\ (\ref{defMmat}).
The point is that the number of nonzero elements of ${\cal M}$ 
is only $O(N^3)$ (not $O(N^4)$).
Indeed, the multiplication ${\cal M} \, v$ can be done economically as follows.

We consider
\beq
w_{a\alpha} ={\cal M}_{a \alpha b \beta} v_{b \beta} \ ,
\eeq
and define the quantities $w '_{a\alpha}$ and $v '_{a \alpha}$, where 
$a$ runs from 1 to $ N^{2}$ as in ${\cal M}'$, by
\beqa
\label{vprime}
v '_{a\alpha} &=& v _{a\alpha}  ~~~~~~~\mbox{for}~~~a=1,\cdots,N^2-1 \ , \n
v '_{N^2 \alpha} &=& - \sum _{i_a = j_a} v _{a\alpha} \ , \\
w' _{a\alpha} &=& {\cal M}'_{a \alpha b \beta}  v ' _{b \beta} \ .
\label{Mbprime}
\eeqa
Now $w_{a\alpha}$ can be written as
\beq
w _{a\alpha} =
\left\{
\begin{array}{ll}
 w' _{a\alpha} - w'_{N^2 \alpha} & \mbox{for}~~~i_a =j_a \\
 w' _{a\alpha} & \mbox{otherwise.}
\end{array}
\right.
\eeq
Thus the problem reduces to calculating the matrix-vector product in
eq.\ (\ref{Mbprime}).
Using definition (\ref{Mprimedef}), we obtain
\beq
(w'_{\alpha })_{ij} 
= (\Gamma ^\mu)_{\alpha \beta} [ A_\mu , v' _{\beta} ]_{ji} ,
\label{bprime}
\eeq
where $w'_{\alpha}$ and $v ' _{\alpha}$ are $N \times N$ matrices
associated with $w'_{a \alpha }$
and $v ' _{a \alpha}$, respectively, as in eq.\ (\ref{texpand}).
The commutator in eq.\ (\ref{bprime}) requires
$O(N^3)$ arithmetic operations. Thus we save $O(N)$ operations.
In addition, we do not have to store neither $g_{abc}$ nor
${\cal M}$. Multiplication of ${\cal M}^\dag$ 
with some vector $v$
is done in the same way.

A similar technique should be used to 
calculate $R_{a\mu}$ in eq.\ (\ref{Rdef}).
Note first that it can be written as
$ R_{c\mu} = T_{c\mu} + (T_{\bar{c}\mu} )^* $,
where $T_{c\mu}$ is given by
\beqa
T_{c\mu}  &=&
\Psi _{a \alpha} 
\left(  \frac{\del {\cal M}}{\del A_{c\mu} } 
\right)_{a \alpha b \beta} \Phi _{b \beta} \ , \nonumber \\
\Psi _{a \alpha } &=& 
({\cal M} _{a\alpha b \beta} \Phi _{b \beta})^{*} \ .
\eeqa
We define $\Phi '$ and $\Psi '$ in terms of $\Phi $ and $\Psi $,
as we defined $v '$ in terms of $v$ before in eq.\ (\ref{vprime}).
Now we can re-write $T_{c\mu}$ as
\beq
T_{c \mu}  =
 \frac{\del }{\del A_{c\mu}} 
\left( \Psi   _{a \alpha } ^{ '} 
({\cal M}  ' )_{a\alpha b \beta} \Phi '  _{b \beta}  \right) .
\eeq
Using again eq.\ (\ref{Mprimedef}), we obtain
\beq
(T_{\mu})_{ij} =
- (\Gamma_{\mu})_{\alpha\beta} 
[\Psi  _{\alpha}^{ '} , \Phi '  _{\beta} ]_{ji} \ ,
\eeq
where $\Phi'_{\alpha}$, $\Psi ' _{\alpha}$ and $T ' _{\mu}$ 
are $N \times N$ matrices associated with
$\Phi  _{a\alpha}^{' }$,
$\Psi  _{a\alpha}^{' } $ and 
$T  _{a\mu}^{'}$, respectively, as in eq.\ (\ref{texpand}).

There are two parameters $\nu$ and $\Delta \tau$
in this algorithm.
We can choose $\nu \Delta \tau$ so that a typical autocorrelation 
time is minimized. We have taken $\nu \Delta \tau =1$ throughout 
the present work, and
$\nu = 200,\, 280,\, 500$ for each of the cases
$N=16,\, 24,\, 32$,
and $\nu = 500,\, 600$ for $N=48$.
Except in Fig.\ \ref{fig:Rnew},
we observed that the results are reasonably well converged
at $\nu = 500$, $\Delta \tau = 0.002$,
so we just present those results.
For Fig.\ \ref{fig:Rnew} we carried out an extrapolation to 
$\Delta \tau =0$ by assuming the $\Delta \tau$
dependence of some observables $Q(\Delta \tau)$ to be
\beq
Q(\Delta \tau) - Q(\Delta \tau=0) 
\sim (\Delta \tau)^2 \cdot
\langle \tr (A_\mu^{~2}) \rangle _{\Delta \tau} \ .
\label{extrapol_assumption}
\eeq
This assumption has been checked for $\langle \tr F^{2}\rangle$ with 
the exact result (\ref{trF2}).
We also observed that
$\langle \tr (A_\mu^{~2}) \rangle _{\Delta \tau}$ behaves as
\beq
\langle \tr (A_\mu^{~2}) \rangle _{\Delta \tau}
\sim
c_1 - c_2  \log \Delta \tau \ ,
\label{trA2div}
\eeq
for small $\Delta \tau$, where 
$c_1$ and $c_2$ are constants depending on $N$.
\footnote{In QCD the $\Delta \tau$ dependence of the systematic error is
$O(\Delta \tau ^2)$ \cite{hybridR}.
A similar argument leads to the assumption
(\ref{extrapol_assumption}).
Due to eq.\ (\ref{trA2div}), the $\Delta \tau$ dependence of 
the systematic error in our case is expected to be
$(\Delta \tau)^2 \log \Delta \tau$.}
This implies that it diverges logarithmically for 
$\Delta \tau \rightarrow 0$, which is consistent with the theoretical
prediction discussed below eq.\ (\ref{Rnewdef}).

Let us comment on the required computational effort
of our algorithm.
The dominant part comes from solving the linear system
(\ref{invD}) using the conjugate gradient method.
First of all, we find that
the number of iterations necessary for the convergence of the method
seems to grow linearly with the size of the matrix ${\cal D}$, namely as $O(N^2)$.
This is much worse than the full QCD case with a fixed quark mass,
where the number of iterations does not depend on the system size.
We may interpret this phenomenon as a sort of ``critical slowing down'', 
since the present system corresponds to QCD in the chiral limit.
As we have seen, the number of arithmetic operations for each iteration
is of order $N^3$.
Therefore, the required computational effort of our algorithm
is estimated to be $O(N^5)$.

For the bosonic case, we use the heat bath algorithm in the way
proposed in Ref.\ \cite{HNT}, which requires an effort of $O(N^4)$.
We note, however, that application of a
Hybrid Monte Carlo algorithm \cite{HMC} allows for
an $O(N^3)$ algorithm for the bosonic case, 
which might be useful for proceeding to much larger $N$.

Finally, we comment on the 
numbers of configurations used for the measurements.
For the supersymmetric case,
they are 3060, 1508, 1296, 436 for $N=16,24,32,48$, respectively.
For the bosonic case, we used 1000 configurations for each $N$.


\section{Results for the bosonic case}

\begin{figure}[htbp]
  \begin{center}
    \includegraphics[height=9cm]{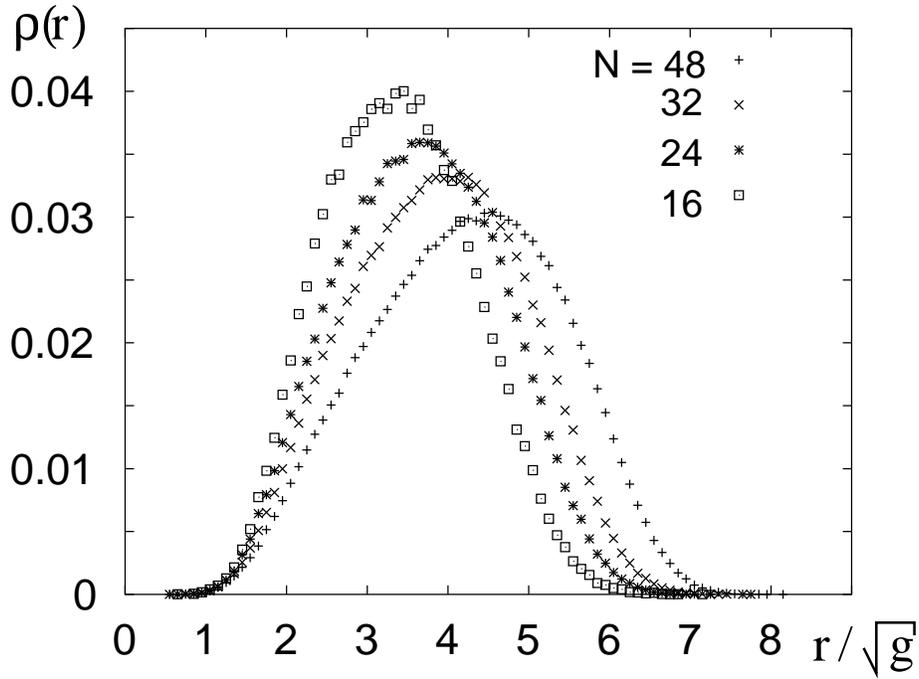}
    \caption{The bosonic distribution of distances $\rho(r)$,
     plotted against $r/\protect\sqrt{g}$ for $N=16$, 24, 32 and 48.}
\label{fig:BEV1}
  \end{center}
\end{figure}

\begin{figure}[htbp]
  \begin{center}
    \includegraphics[height=12cm,angle=270]{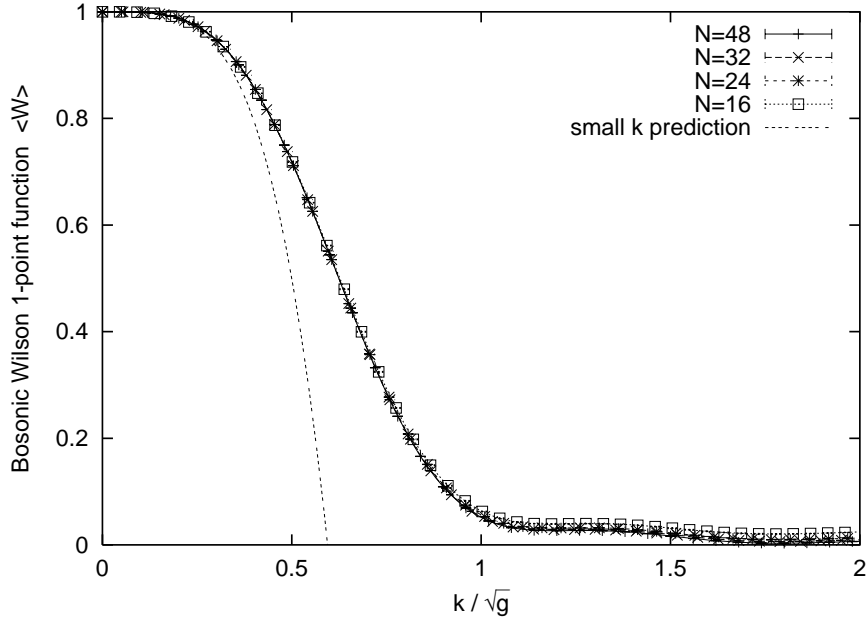}
    \caption{The bosonic Wilson 1-point function $\langle W \rangle$, 
    plotted against $k/\protect\sqrt{g}$. In this case, the small
    $k$ prediction amounts to $1-(N/6) k^{4}$.}
\label{fig:Wilson1-bos}
  \end{center}
\end{figure}

\begin{figure}[htbp]
  \begin{center}
    \includegraphics[height=12cm,angle=270]{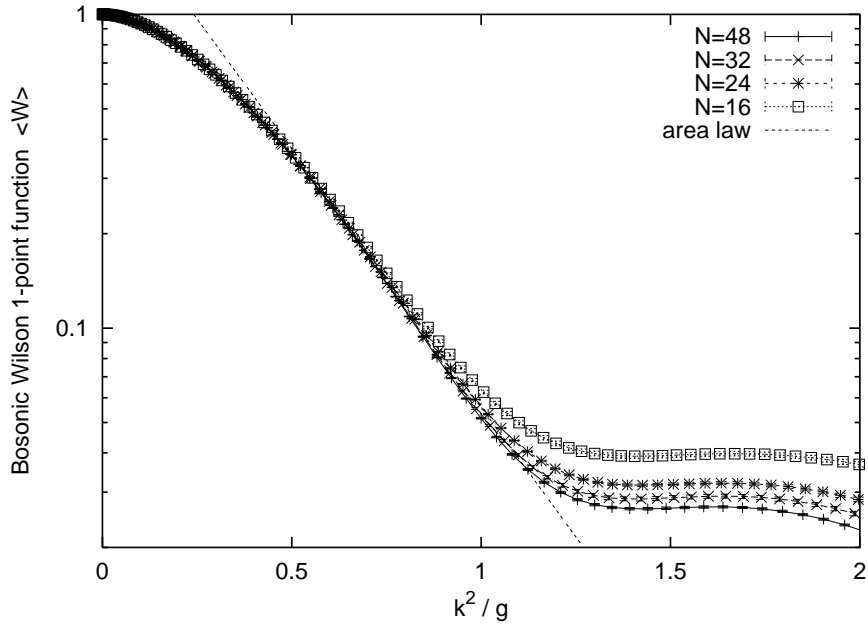}
    \caption{The bosonic Wilson 1-point function $\langle W \rangle$
             is plotted now logarithmically
             against $k^{2}/g$, in order to visualize the extent of 
             the area law behavior.}
\label{fig:Wilson1-log-bos}
  \end{center}
\end{figure}

\begin{figure}[htbp]
  \begin{center}
    \includegraphics[height=12cm,angle=270]{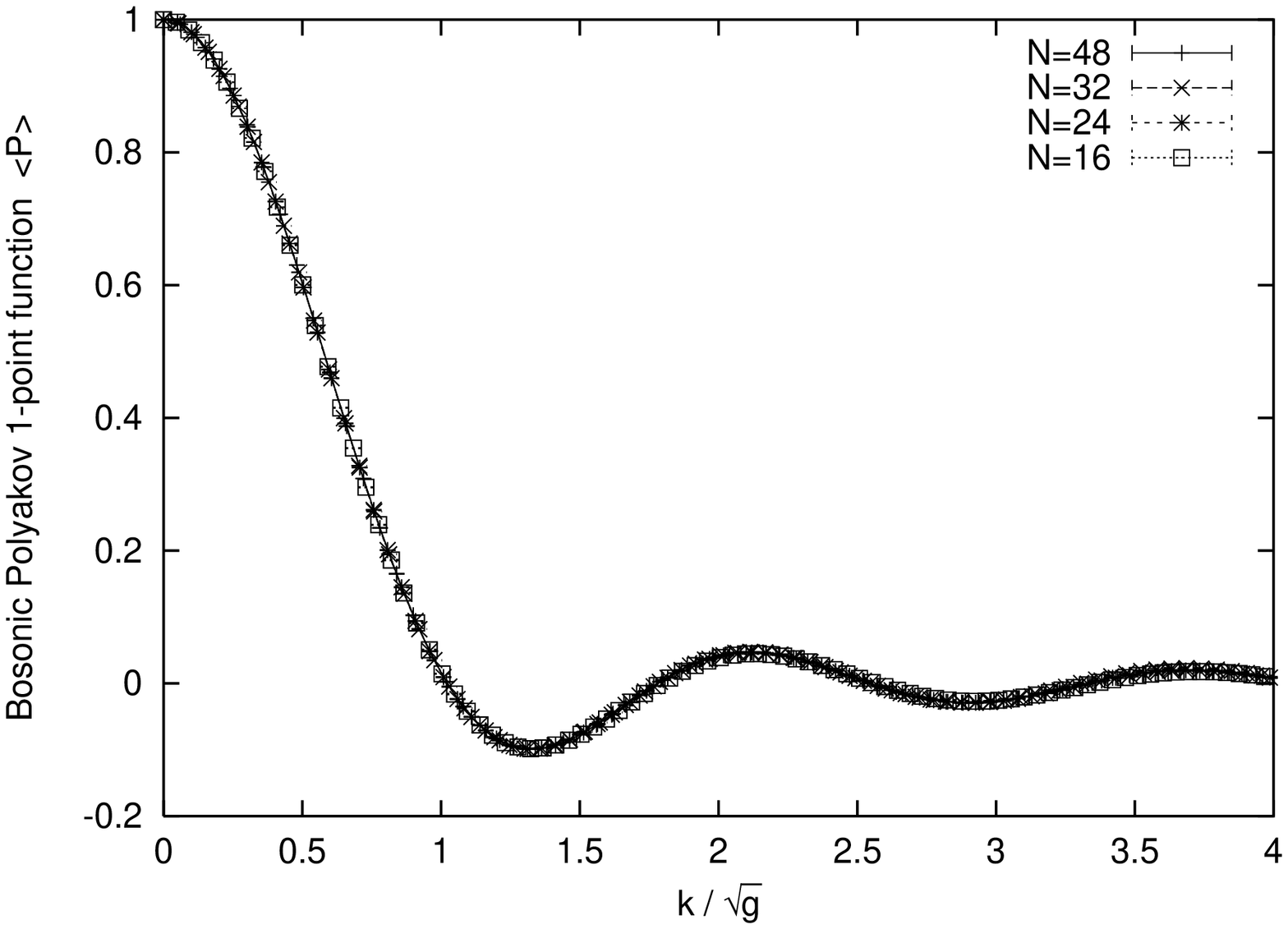}
    \caption{The bosonic Polyakov 1-point function $\langle P \rangle$, 
             plotted  against $k/\protect\sqrt{g}$.}
\label{fig:Polyakov1-bos}
  \end{center}
\end{figure}

\begin{figure}[htbp]
  \begin{center}
    \includegraphics[height=12cm,angle=270]{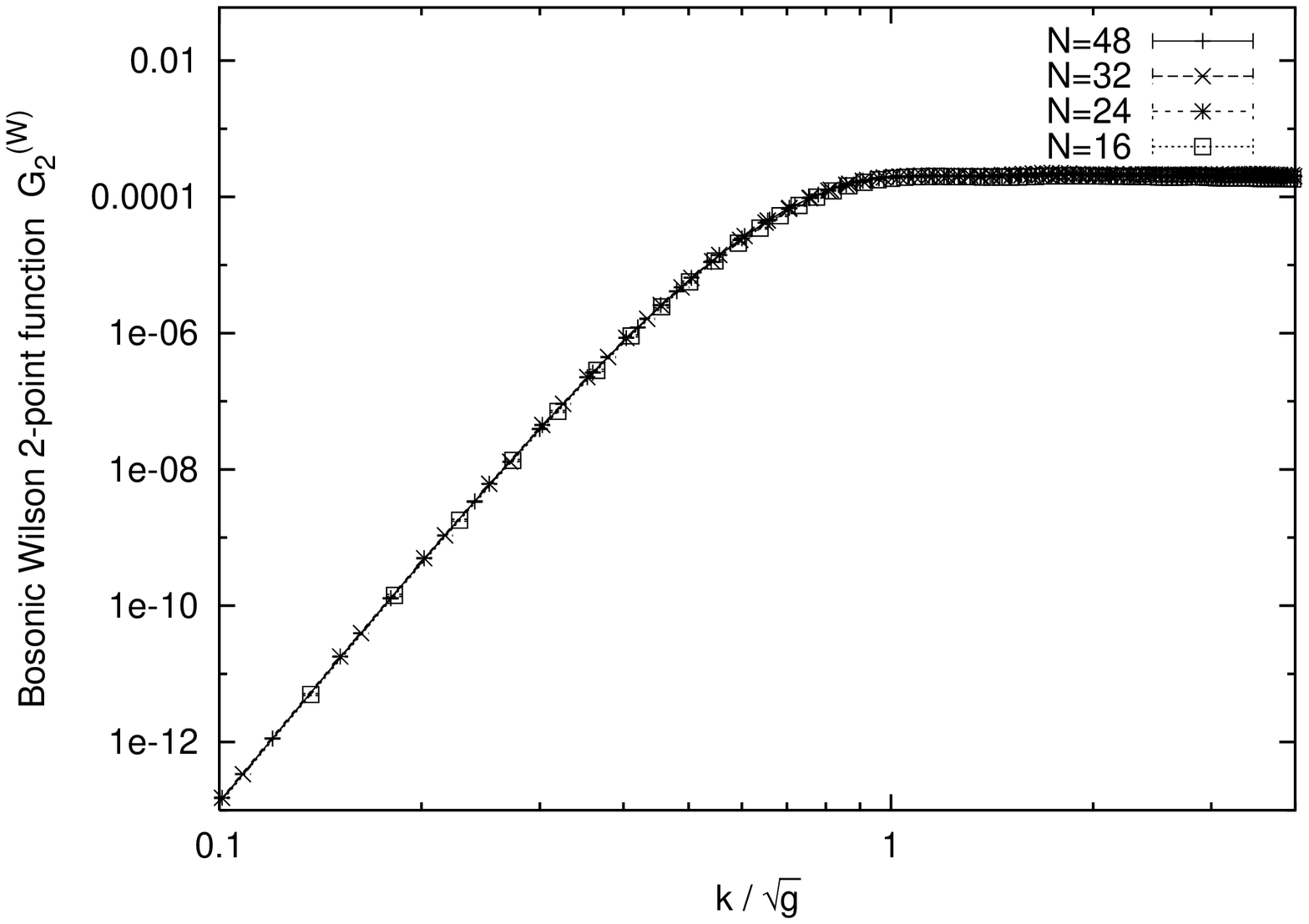}
    \caption{The bosonic Wilson 2-point function $G_{2}^{(W)}$,
 multiplied by $Z^{2} \propto N^2$, 
 plotted against $k/\protect\sqrt{g}$.}
\label{fig:G2-bos}
  \end{center}
\end{figure}

\begin{figure}[htbp]
  \begin{center}
    \includegraphics[height=12cm,angle=270]{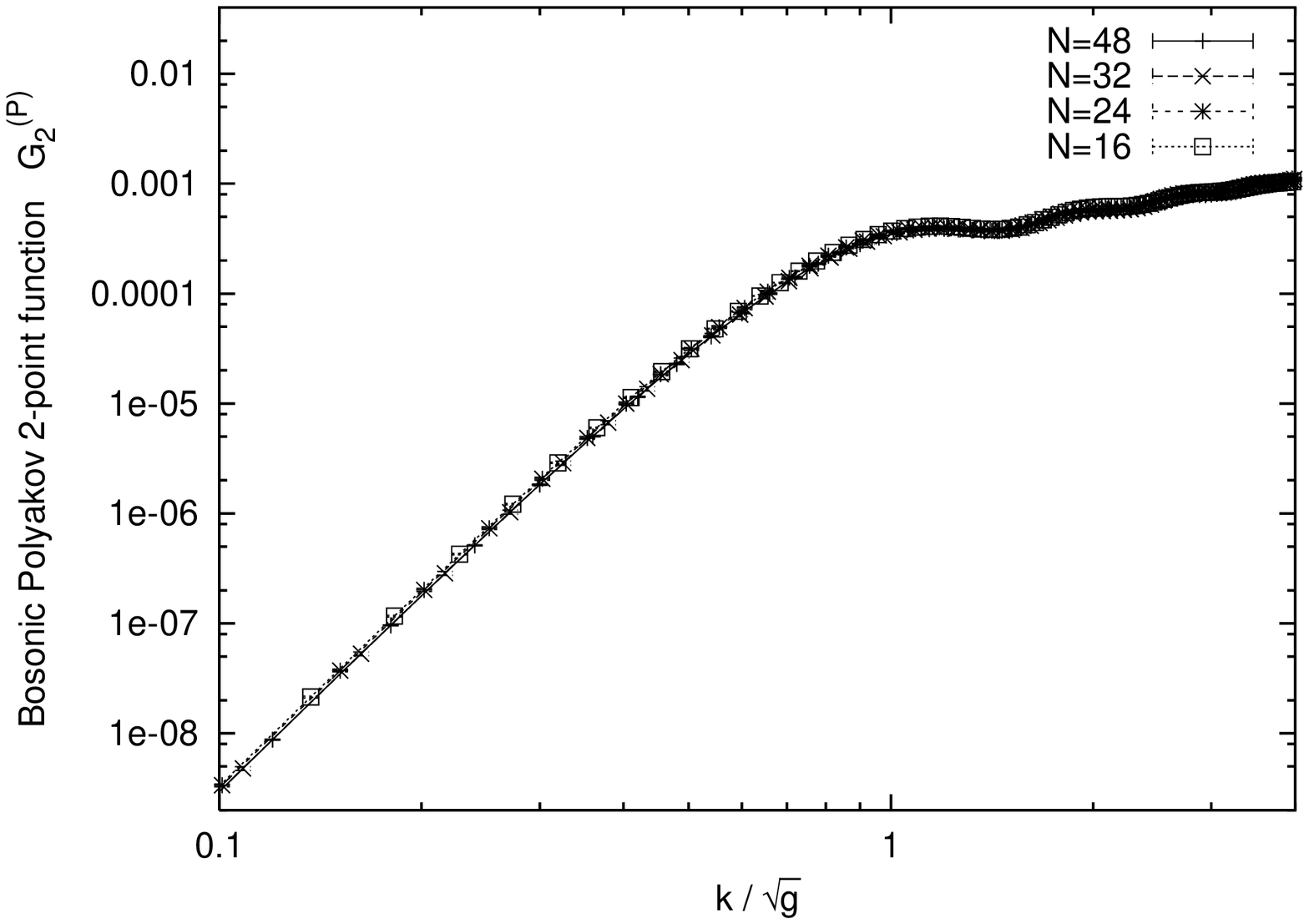}
    \caption{The bosonic Polyakov 2-point function $G_{2}^{(P)}$, 
 multiplied by $Z^{2} \propto N^2$, 
 plotted against $k/\protect\sqrt{g}$.}
\label{fig:P2-bos}
  \end{center}
\end{figure}

\begin{figure}[hbt]
   \begin{tabular}{cc}
      \hspace{-0.6cm}
\def\fpsangle{270} \epsfysize=85mm \fpsbox{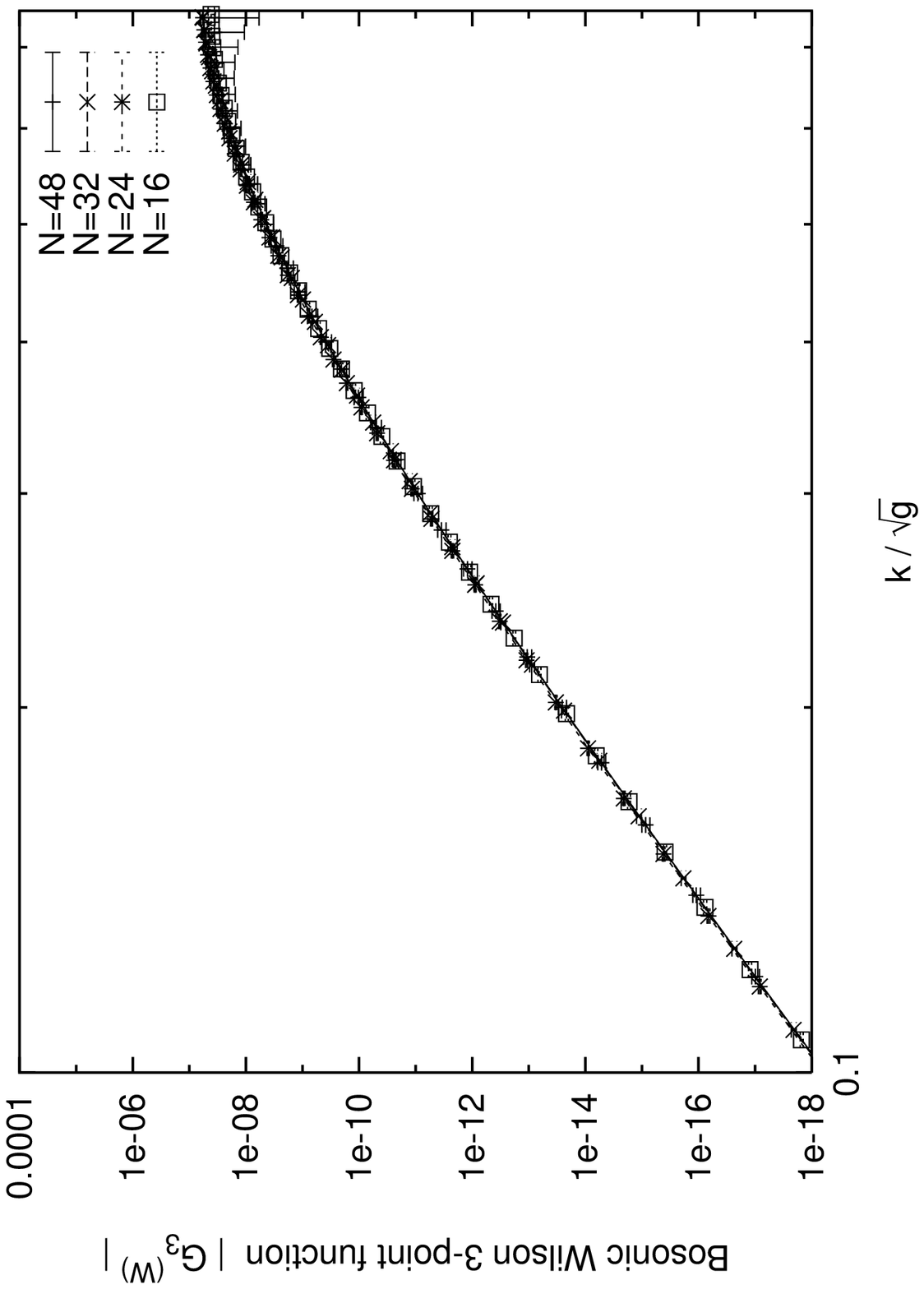} &
      \hspace{-0.6cm}
\def\fpsangle{270} \epsfysize=85mm \fpsbox{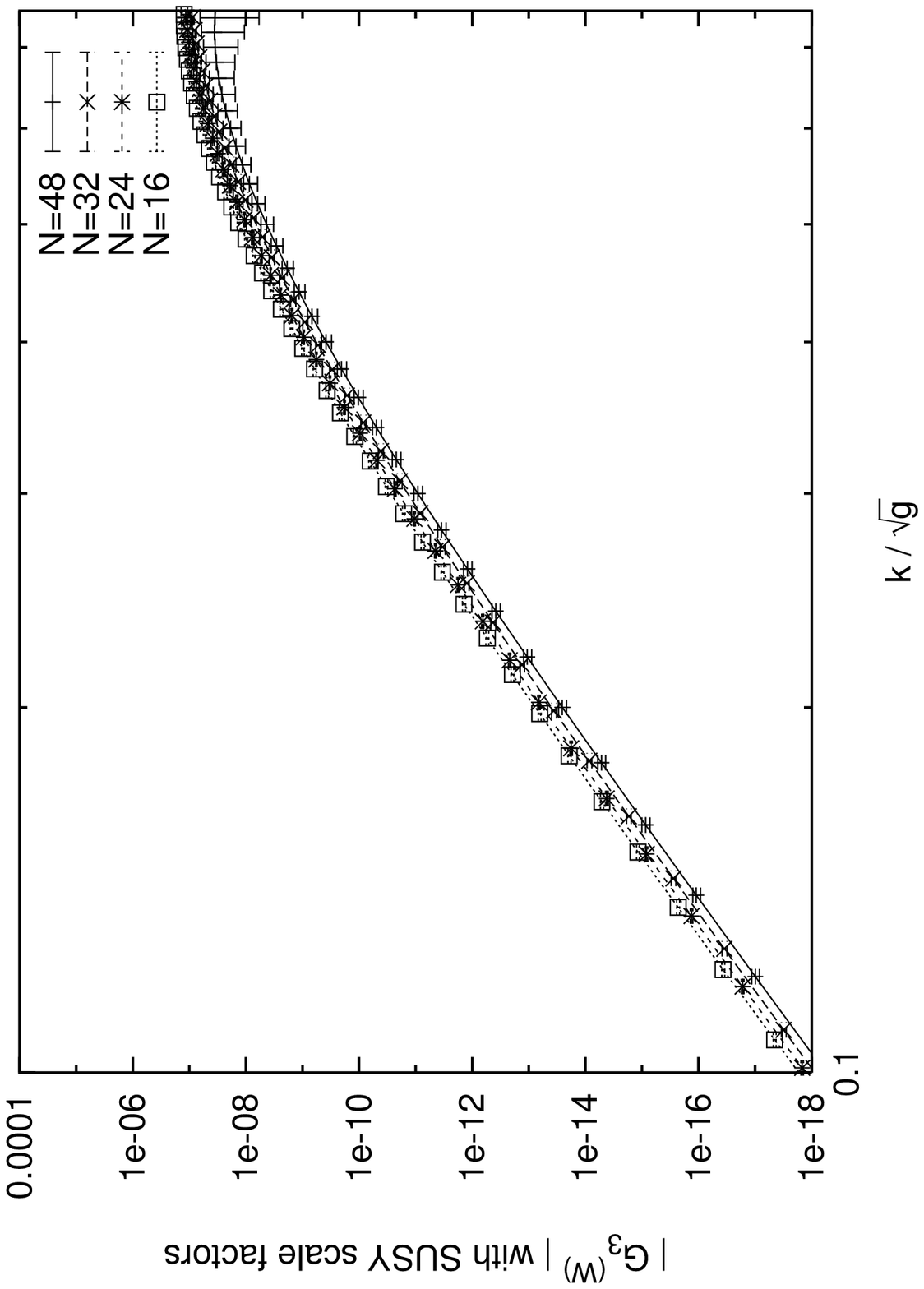}
   \end{tabular}
\caption{The bosonic Wilson 3-point function $G_3^{(W)}$,
 multiplied by $Z^{3} \propto N^4$, plotted against $k/\protect\sqrt{g}$
on the left. On the right we show the corresponding plot using
the SUSY prediction $Z^{3} \propto N^{3}$ instead, which leads to
an inferior level of scaling.}
\label{fig:wil3-bos}
\end{figure}

\begin{figure}[hbt]
   \begin{tabular}{cc}
      \hspace{-0.5cm}
\def\fpsangle{270} \epsfysize=84mm \fpsbox{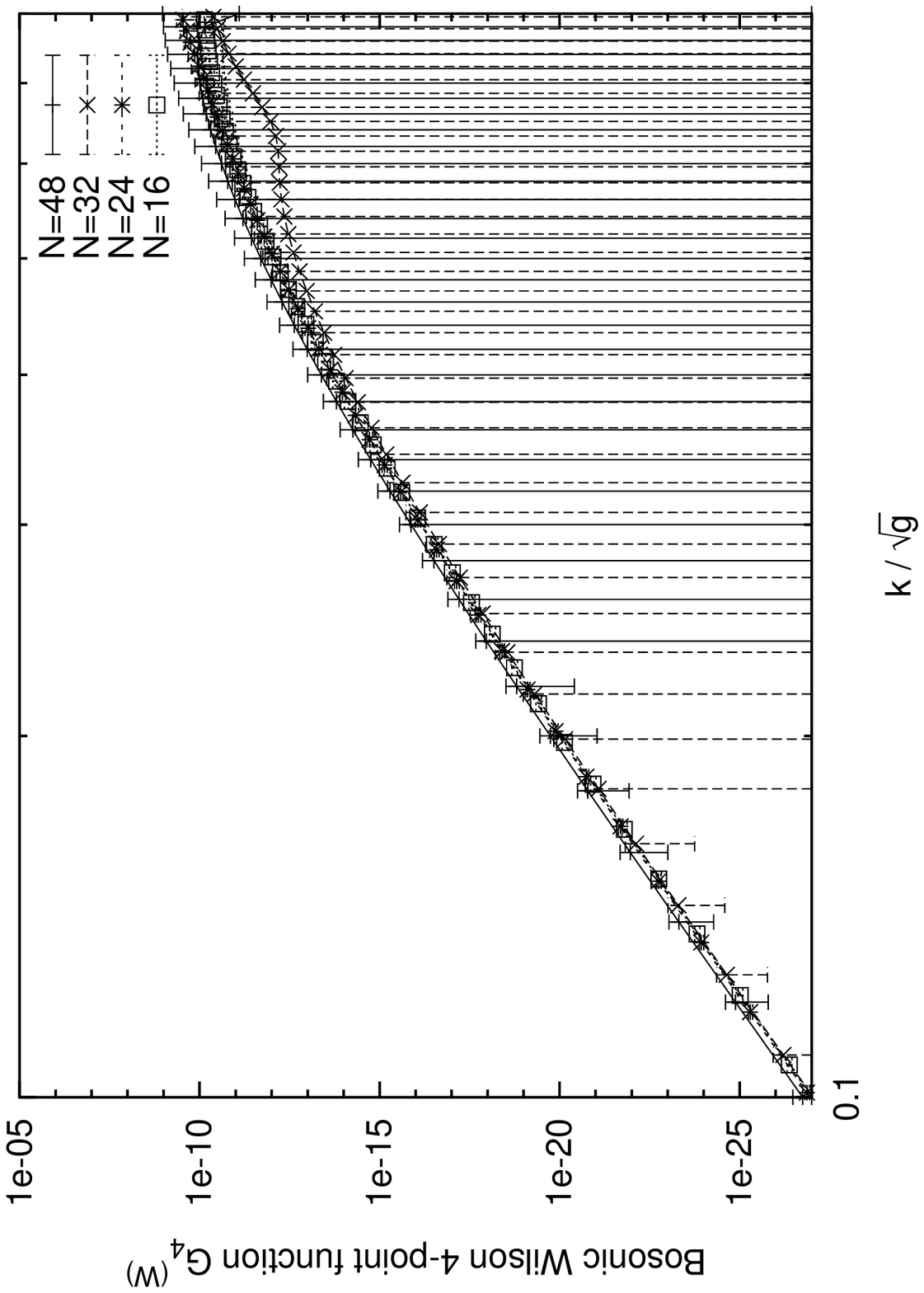} &
      \hspace{-0.55cm}
\def\fpsangle{270} \epsfysize=84mm \fpsbox{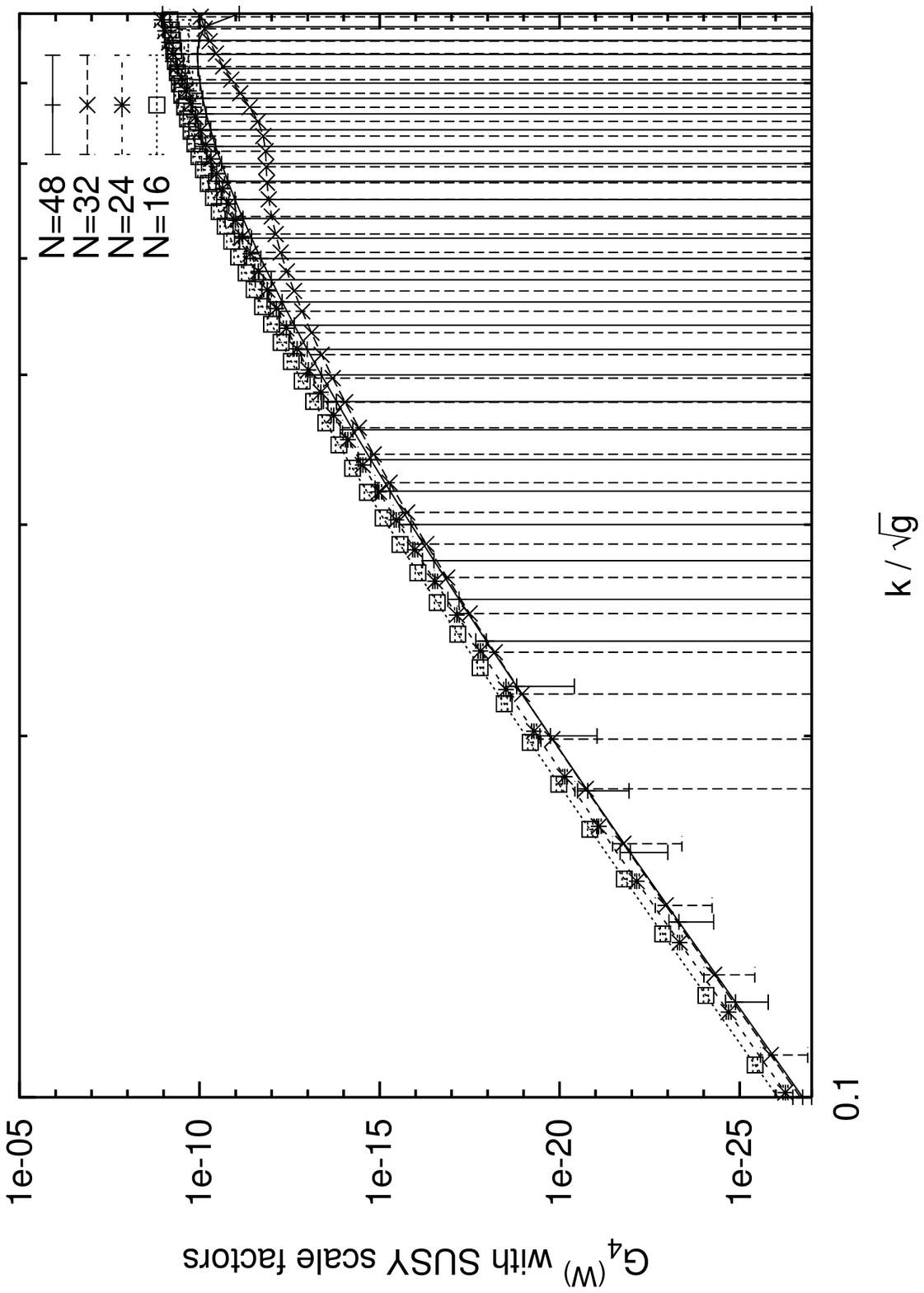}
   \end{tabular}
\caption{The bosonic Wilson 4-point function $G_{4}^{(W)}$,
 multiplied by $Z^{4} \propto N^6$,
 plotted against $k/\protect\sqrt{g}$ on the left.
 On the right we show the corresponding plot using
the SUSY prediction $Z^{4} \propto N^{4}$ instead, which leads to
an inferior level of scaling.}
\label{fig:wil4-bos}
\end{figure}

\setcounter{equation}{0}
\label{bosonic_results}
\renewcommand{\theequation}{A.\arabic{equation}}
\hspace*{\parindent}
For comparison we show in this appendix the results
for the bosonic case.
By the bosonic case we mean a model obtained
by just dropping the fermions from the supersymmetric matrix model
described by the partition function (\ref{action}).
Fig.\ \ref{fig:BEV1} 
shows the distribution $\rho (r)$ defined in Section \ref{spacetime}.
Figs.\ \ref{fig:Wilson1-bos} 
to \ref{fig:wil4-bos} 
show the Wilson loop and Polyakov line correlators defined
in Section \ref{wilsonloops}.
We take $g\propto 1/\sqrt{N}\,$ \ ($g=1$ \ for \ $N=48$)
and plot the results against $k_{\phys} = k/\sqrt{g}$,
as in the supersymmetric case.
We multiply the results
by $(N/48)^{2(n-1)}$ for $n$-point functions.

The data scale nicely in agreement with
the theoretical prediction for large $N$ given by
eq.\ (\ref{bosonicscaling}).
For comparison we also show the 3-point and the 4-point
Wilson loop correlators with the renormalization factors,
which were used successfully in Sec.\ 4 for the supersymmetric
case. We see very clearly that the bosonic prediction is
the correct one in this case.

\end{document}